\UseRawInputEncoding
\documentclass[11pt,nofootinbib,preprint]{revtex4}
\usepackage{mathrsfs}
\usepackage{graphicx}
\usepackage{amsmath}
\usepackage{amsfonts}
\usepackage{amssymb}
\usepackage{color}

\usepackage{epsfig}
\usepackage{CJK}
\usepackage{graphicx}
\usepackage{epsfig}
\usepackage{eepic}
\usepackage{bbm}
\usepackage{dcolumn}% Align table columns on decimal point
\usepackage{bm}% bold math
\usepackage{ulem}
\usepackage{slashbox}
\usepackage{multirow}% for the tabular, added in 30.01.2013
\usepackage{slashed}

\newcommand{\omits}[1]{}

\def\bc{\begin{center}}

\def\ec{\end{center}}
\def\be{\begin{eqnarray}}
\def\ee{\end{eqnarray}}

\definecolor{dyellow}{rgb}{1.,0.8,.0}
\definecolor{myblue}{rgb}{.1,.1,.7}
\definecolor{dcyan}{rgb}{.0,.6,.6}
\definecolor{cyan}{rgb}{0.4,1.0,1.0}
\definecolor{dmagenta}{rgb}{0.6,0.0,0.6}
\definecolor{brown}{rgb}{0.6,0.2,0.}
\definecolor{darkblue}{rgb}{.0,.0,0.5}
\definecolor{darkred}{rgb}{0.75,0.0,0.0}
\definecolor{orange}{rgb}{1.,.6,.0}
\definecolor{dorange}{rgb}{0.8,.4,.0}
\definecolor{green}{rgb}{0.0,1.0,0.0}
\definecolor{darkgreen}{rgb}{0.0,0.6,0.0}
\definecolor{purple}{rgb}{.4,.0,.4}
\definecolor{lightgrey}{rgb}{0.7, 0.7, 0.7}
\definecolor{grey}{rgb}{0.4, 0.4, 0.4}
\newcommand{\nc}{\newcommand}
\nc{\rnc}{\renewcommand} \nc{\ket}[1]{\left | \, #1 \right \rangle}
\nc{\bra}[1]{\left \langle #1 \, \right |}
\nc{\ua}{\uparrow} \nc{\da}{\downarrow}

\nc{\braket}[2]{\langle\, #1\,|\,#2\,\rangle}
\nc{\half}{\frac{1}{2}}

\nc{\prj}{\mathcal{P}} \nc{\hilb}{\mathcal{H}}
\nc{\pth}{\mathcal{C}} \nc{\inprod}[2]{\braket{#1}{#2}}
\nc{\upket}{\ket{\uparrow}} \nc{\downket}{\ket{\downarrow}}
\nc{\upbra}{\bra{\uparrow}} \nc{\downbra}{\bra{\downarrow}}

\begin{document}

%\preprint{hep-th/yymmnnn}

\title{Deriving the PEE proposal from the Locking bit thread configuration}

\author{Yi-Yu Lin$^1$} \email{linyy27@mail2.sysu.edu.cn}
\author{Jia-Rui Sun$^{1}$} \email{sunjiarui@mail.sysu.edu.cn}
\author{Jun Zhang$^{1}$} \email{zhangj626@mail2.sysu.edu.cn}

\affiliation{${}^1$School of Physics and Astronomy, Sun Yat-Sen University, Guangzhou 510275, China}
%\affiliation{${}^2$State Key Laboratory of Theoretical Physics,, Institute of Theoretical Physics, Chinese Academy of Sciences, Beijing 100190, China}
%\affiliation{${}^2$Department of Physics and Center for Mathematics and Theoretical Physics, National Central University, Chungli 320, Taiwan}
%\date{\today}

%% REVTEX4
%\maketitle

\begin{abstract}
In the holographic framework, we argue that the partial entanglement entropy (PEE) can be explicitly interpreted as the component flow flux in a locking bit thread configuration. By applying the locking theorem of bit threads, and constructing a concrete locking scheme, we obtain a set of uniquely determined component flow fluxes from this viewpoint, and successfully derive the PEE proposal and its generalized version in the multipartite cases. Moreover, from this perspective of bit threads, we also present a coherent explanation for the coincidence between the BPE (balanced partial entanglement)/EWCS (entanglement wedge cross section) duality proposed recently and the EoP (entanglement of purification)/EWCS duality. We also discuss the issues implied by this coincident between the idea of the PEE and the picture of locking thread configuration.

\end{abstract}

%% REVTEX4

\maketitle
\tableofcontents

%%%%%%%%%%%%%%%%%%%%%%%%%%%%%%%%%%%%%%%%%%%%%%%%%%%%%%%%%%%%%%%%%%%%%%
\section{Introduction}
%%%%%%%%%%%%%%%%%%%%%%%%%%%%%%%%%%%%%%%%%%%%%%%%%%%%%%%%%%%%%%%%%%%%%%
Quantum entanglement is one of the most intriguing fundamental features of quantum mechanics. In recent years, in the framework of holographic principle~\cite{Maldacena:1997re,Gubser:1998bc,Witten:1998qj}, inspired by the famous RT formula for calculating entanglement entropy~\cite{Ryu:2006bv,Ryu:2006ef,Hubeny:2007xt}, quantum entanglement has been widely conjectured (or believed) as the key ingredient for the emergence of bulk spacetime~\cite{VanRaamsdonk:2010pw,Lashkari:2013koa,Faulkner:2013ica,Faulkner:2017tkh,Bao:2019bib,Lin:2020thc,Lin:2020,Sun:2019ycv,Bao:2018pvs,Swingle:2009bg,Swingle:2012wq,Milsted:2018san,Pastawski:2015qua,Hayden:2016cfa,Qi:2013caa,Agon:2020mvu,Lin:2020yzf}. For a system in a pure state, the standard measure of entanglement is the entanglement entropy $S\left( A \right)$, which is a highly nonlocal quantity describing the entanglement between some subsystem $A$ and its complement ${A_c}$. Naturally, it is tempting to express this quantity in a more refined way as the sum of the contributions of each local degree of freedom in $A$. Indeed, the concept of the entanglement contour is an explicit realization of this idea~\cite{vidal2014}. Briefly, the entanglement contour ${{f_A}\left( x \right)}$ is a density function of entanglement entropy $S\left( A \right)$, satisfying 
\be S\left( A \right) = \int_A {{f_A}\left( x \right)dx} ,\ee
where $x$ represents the spatial coordinates of region $A$. Technically, it is more tractable to study the partial entanglement entropy (PEE) ${s_A}\left( {{A_i}} \right)$ of some finite size subset ${{A_i}}$ of $A$~\cite{vidal2014,Wen:2018whg,Kudler-Flam:2019oru,Wen:2019ubu,Wen:2019iyq}, which is defined as 
\be{s_A}\left( {{A_i}} \right) \equiv \int_{{A_i}} {{f_A}\left( x \right)dx} .\ee
In other words, the PEE ${s_A}\left( {{A_i}} \right)$ captures the contribution from ${{A_i}}$ to entanglement entropy $S\left( A \right)$. The concepts of the PEE and the entanglement contour not only have a range of applications in studying the entanglement structures in condensed matter theory~\cite{vidal2014,Kudler-Flam:2019oru,DiGiulio:2019lpb,MacCormack:2020auw}, but also have enlightening significance in the holographic framework~\cite{Wen:2018whg,Wen:2019ubu,Wen:2018mev}. However, so far the fundamental definition of the PEE based on the reduced density matrix has not been established. Rather, it is required to satisfy a series of reasonable conditions according to its physical meaning~\cite{vidal2014,Wen:2019iyq}:

1. Additivity: decomposing ${{A_i}}$ as ${A_i^1}$ and ${A_i^2}$, by definition we should have
\be{s_A}\left( {{A_i}} \right) = {s_A}\left( {A_i^1} \right) + {s_A}\left( {A_i^2} \right) .\ee

2. Invariance under local unitary transformations: ${s_A}\left( {{A_i}} \right)$ should be invariant under any local unitary transformations inside ${{A_i}}$ or ${A_c}$.

3. Symmetry: for any symmetry transformation $T$ under which $TA = A'$ and $T{A_i} = {A_i}'$, we have ${s_A}\left( {{A_i}} \right) = {s_{A'}}\left( {{A_i}'} \right)$.

4. Normalization: $S\left( A \right) = {\left. {{s_A}\left( {{A_i}} \right)} \right|_{{A_i} \to A}}$.

5. Positivity: ${s_A}\left( {{A_i}} \right) \ge 0$.

6. Upper bound: ${s_A}\left( {{A_i}} \right) \le S\left( A \right)$.

7. Symmetry under the permutation: since the PEE ${s_A}\left( {{A_i}} \right)$ captures the correlation between the subset ${A_i}$ and ${A_c}$ in some sense, it should be invariant under the permutation between ${A_i}$ and ${A_c}$~\cite{Wen:2019iyq}. To manifest this permutation symmetry, we can express the PEE in the following way:

\be\label{per}{s_A}\left( {{A_i}} \right) = P\left( {{A_i},{A_c}} \right) = P\left( {{A_c},{A_i}} \right) = {s_{{{\left( {{A_i}} \right)}_c}}}\left( {{A_c}} \right) ,\ee
where ${\left( {{A_i}} \right)_c}$ represents the complement of ${{A_i}}$.

Since the above requirements are not sufficient to uniquely determine the PEE in general, ~\cite{Wen:2018whg,Wen:2019ubu,Kudler-Flam:2019oru} proposed a PEE proposal, which claims that the PEE can be obtained by an additive linear combination of subset entanglement entropies.\footnote{In fact, there are other proposals for the PEE, see e.g.~\cite{vidal2014,Coser:2017dtb,Tonni:2017jom,Kudler-Flam:2019nhr,Wen:2018mev,Wen:2018whg,Wen:2019iyq,Han:2019scu}. Although these proposals have different physical motivations, the PEE calculated by different approaches are highly consistent with each other~\cite{Wen:2018whg,Wen:2019iyq,Wen:2018mev,Han:2019scu,Kudler-Flam:2019nhr}.}

On the other hand, the concept of the entanglement contour and the PEE naturally reminds us of the picture of bit threads. The formulation of bit threads arose from the fact that it can equivalently describe the famous RT formula for entanglement entropy~\cite{Freedman:2016zud,Cui:2018dyq}\footnote{See also another interesting reformulation of the RT prescription in terms of the so-called calibrations in~\cite{Bakhmatov:2017ihw}.}. Since it is endowed with an intuitive picture, in a sense, it can also be regarded as providing a graphical explanation for the RT formula. Bit threads are a kind of unoriented bulk curves which is required to end on the boundary, but can travel through the bulk spacetime. In addition, they are required to satisfy certain constraints. They are required to be divergenceless, and subject to the rule that the thread density is less than 1 everywhere (in units where $4{G_ N} = 1$. This convention will always be adopted in this paper). According to the so-called max flow-min cut theorem, it can be shown that the maximal flux of bit threads (over all possible bit thread configurations) through a boundary subregion $A$ is equal to the area of the bulk minimal surface homologous to $A$, i.e., the RT surface $\gamma \left( A \right)$. A thread configuration that can achieve this maximal flux is said to $lock$ $A$. Therefore, the RT formula can be expressed in another way, that is, the entropy of a boundary subregion $A$ is equal to the flux of the locking thread configuration passing through $A$~\cite{Freedman:2016zud,Cui:2018dyq},
\be S\left( A \right) = {Flux}_{\rm locking}\left( A \right) .\ee
One of intriguing ideas of particular interest is to consider how gravity emerges from entanglement using the bit threads language, see the recent work in~\cite{Lin:2020yzf,Agon:2020mvu}. Furthermore, recently a generalized version of bit threads (named ``quantum bit threads'') was proposed in~\cite{Agon:2021tia,Rolph:2021hgz} to include the quantum corrections and discuss the quantum extremal islands in the context of the information loss problem~\cite{Penington:2019npb,Almheiri:2019psf,Almheiri:2019hni,Almheiri:2019qdq}.

Obviously, one can see the tantalizing similarity between the picture of bit threads and the concept of the entanglement contour and the PEE. Furthermore, since the properties of bit threads are designed to mathematically recover the RT formula in such a delicate way, they may provide a more solid foundation for the idea of entanglement contour, at least in the holographic aspect. Actually, there has been a preliminary discussion of the relationship between entanglement contour and bit threads in~\cite{Kudler-Flam:2019oru}, in which the entanglement contour ${f_A}\left( x \right)$ is explicitly interpreted as the $flow$ ${v\left( x \right)}$ describing the bit threads, i.e.,
\be{f_A}\left( x \right) = \left| {v\left( x \right)} \right| .\ee
Moreover, it has been shown that this identification is consistent with the series of conditions of the PEE and the entanglement contour mentioned above.

In this paper, we will use the notion of $multiflow$, which is a more powerful mathematical tool than the $flow$ in the formulation of bit threads, to further sharpen this viewpoint, making it more clearly and adapted to more general situations. In particular, we will show that, in the holographic framework, we can naturally derive the PEE proposal using the language of bit threads. More specifically, we explicitly identify the PEE as the flux of the $component~flow$ in a locking bit thread configuration~\cite{Headrick:2020gyq,Lin:2020yzf}. In other words, we identify the entanglement contour as the component flow in a multiflow that describes the locking thread configuration. On the other hand, we will also show that there exist some subtle problems between the idea of the PEE and the locking problem in the formulation of bit threads, which seems to imply that one
or both of them are limited in some sense. We lay out some discussion on these issues, but the further concrete reconciliation will be left as an open question for the moment.

The structure of this paper is as follows: In section \ref{sec2}, we review the background knowledge about bit threads and locking thread configurations. We first review the basic concepts of bit threads in section \ref{subsec2.1}, then in section \ref{subsec2.2} we review a latest technical development of bit threads, namely the locking program of bit threads, contributed by the recent paper~\cite{Headrick:2020gyq}. In particular, the latter part is the preparatory knowledge on which our study intimately relies, and we summarize all the existence theorems for locking thread configurations that will be used in this paper for convenience. Section \ref{sec3} is the central part of our work. In section \ref{subsec3.1}, we clarify our motivation and explicitly identify the PEE as the flux of the component flow in a locking bit thread configuration. In section \ref{subsec3.2}, we derive the PEE proposal in the case that the subsystem is divided into three parts, based on the locking theorem of bit threads. With this understanding, we then explain the interesting coincidence between the so-called balanced partial entanglement (BPE) and the entanglement of purification (EoP) in section \ref{subsec3.3}. In section \ref{sec4}, we further derive the PEE proposal in the multipartite situations from the picture of bit threads, which results in a conceptual discussion
of bit threads and the PEE. More specifically, in section \ref{subsec4.1}, we construct a concrete locking scheme to show that it is possible to lock enough RT surfaces to have a full rank system of equations between the entanglement entropies of these surfaces and components of the multiflow which connect different regions. Then in section~\ref{subsec4.2}, we discuss the limitations of the locking ability of bit threads at the current stage and the possible solution to this issue. The conclusion and discussion are given in section \ref{sec5}.

%%%%%%%%%%%%%%%%%%%%%%%%%%%%%%%%%%%%%%%%%%%%%%%%%%%%%%%%%%%%%%%%%%%%%%

%%%%%%%%%%%%%%%%%%%%%%%%%%%%%%%%%%%%%%%%%%%%%%%%%%%%%%%%%%%%%%%%%%%%%%
\section{Background review }\label{sec2}

%%%%%%%%%%%%%%%%%%%%%%%%%%%%%%%%%%%%%%%%%%%%%%%%%%%%%%%%%%%%%%%%%%%%%%
%%%%%%%%%%%%%%%%%%%%%%%%%%%%%%%%%%%%%%%%%%%%%%%%%%%%%%%%%%%%%%%%%%%%%%
%%%%%%%%%%%%%%%%%%%%%%%%%%%%%%%%%%%%%%%%%%%%%%%%%%%%%%%%%%%%%%%%%%%%%%

\subsection{The basics of bit threads}\label{subsec2.1}

Bit threads are unoriented bulk curves which end on the boundary and subject to the rule that the thread density is less than 1 everywhere (in units where $4{G_ N} = 1$). In particular, this thread density bound implies that the number of threads passing through the minimal surface $\gamma \left( A \right)$ that separates a boundary subregion $A$ and its complement ${{A_{\rm{c}}}}$ cannot exceed its area $Area\left( {\gamma \left( A \right)} \right)$, hence the flux of bit threads $Flux\left( A \right)$ connecting $A$ and its complement ${{A_{\rm{c}}}}$ does not exceed $Area\left( {\gamma \left( A \right)} \right)$:
\be\label{bound} Flux\left( A \right) \le Area\left( {\gamma \left( A \right)} \right) .\ee
Borrowing terminology from the theory of flows on networks, a thread configuration is said to $lock$ the region $A$ when the bound~(\ref{bound}) is saturated. Actually, this bound is tight: for any $A$, there does exist a locking thread configuration satisfying:
\be Flu{x_{{\rm{locking}}}}\left( A \right) = Area\left( {\gamma \left( A \right)} \right) .\ee
This theorem is known as max flow-min cut theorem (see \cite{Headrick:2017ucz} and references therein), that is, the maximal flux of bit threads (over all possible bit thread configurations) through a boundary subregion $A$ is equal to the area of the bulk minimal surface $\gamma \left( A \right)$ homologous to $A$. Therefore, the famous RT formula which relates the entanglement entropy of a boundary subregion $A$ and the area of the bulk minimal extremal surface ${\gamma \left( A \right)}$ homologous to A:
\be S\left( A \right) = Area\left( {\gamma \left( A \right)} \right) ,\ee
can be expressed in another way, that is, the entropy of a boundary subregion $A$ is equal to the flux of the locking thread configuration passing through $A$:
\be S\left( A \right) = Flu{x_{{\rm{locking}}}}\left( A \right) .\ee

When the bit threads are required to be locally parallel, one can use the language of $flow$ to describe the behavior of bit threads conveniently in mathematics, that is, using a vector field $\vec v$ to describe the bit threads, just as using the magnetic field $\vec B$ to describe the magnetic field lines. The difference is that for the latter we regard the magnetic field itself as the more fundamental concept, while for the former we consider the threads to be more fundamental. The constraints on the bit threads can then be expressed as the requirements for the flow $\vec v$ as follows,
\be\nabla  \cdot \vec v &=& 0,\\
\rho \left( {\vec v} \right) &\equiv & \left| {\vec v} \right| \le 1.\ee

For situations involving more than one pair of boundary subregions, the concept of $thread~ bundles$ is also useful. The threads in each thread bundle are required to connect only a specified pair of boundary subregions, while still satisfy the constraints of bit threads. Specifically, one can use a set of vector fields ${\vec v_{ij}}$ to represent each thread bundle connecting the ${A_i}$ region and ${A_j}$ region respectively. The set $V$ of vector fields ${\vec v_{ij}}$ is referred to as a $multiflow$, and each ${\vec v_{ij}}$ is called a $component~ flow$, satisfying (Note that in the present paper we will define ${\vec v_{ij}}$ only with $i < j$ for convenience, which is slightly different from (but equivalent) convention adopted in~\cite{Cui:2018dyq}, where the fields ${\vec v_{ij}}$ were also defined for $i \ge j$, but with the constraint ${{\vec v}_{ji}} =  - {{\vec v}_{ij}}$)
\be\nabla  \cdot {\vec v_{ij}} &=& 0,\\
\rho (V) &\le & 1,\\
\hat{n}\cdot \vec v_{ij}|_{A_k} &=& 0,\quad({\rm for}\quad k \ne i,j) .\ee
It is worth noting that, since in the situation of multiflows, the threads are not necessarily locally parallel, there are various natural ways the density can be defined, and therefore bounded. As we will see in the next subsection, different definitions of the thread density will actually affect the ability of a thread configuration to lock a set of boundary regions.

%%%%%%%%%%%%%%%%%%%%%%%%%%%%%%%%%%%%%%%%%%%%%%%%%%%%%%%%%%%%%%%%%%%%%%
\subsection{ Locking theorems of bit threads}\label{subsec2.2}
%%%%%%%%%%%%%%%%%%%%%%%%%%%%%%%%%%%%%%%%%%%%%%%%%%%%%%%%%%%%%%%%%%%%%%
%%%%%%%%%%%%%%%%%%%%%%%%%%%%%%%%%%%%%%%%%%%%%%%%%%%%%%%%%%%%%%%%%%%%%%
For a single boundary subregion $A$, the max flow-min cut theorem directly indicates that one can find a thread configuration that can lock the specified boundary subregion (and its complement simultaneously). In other words, there exist thread configurations that can lock a set of boundary regions $I = \left\{ {{A_1},{A_{\rm{c}}}} \right\}$, and there is typically an infinite number of choices. However, one can further ask, can we find a locking thread configuration that can lock an arbitrary specified set of subregions simultaneously? The question becomes very nontrivial. Broadly speaking, it depends not only on the relative space position relations between these specified subregions, but also on the properties we assign to the bit threads, in particular, the precise definition of the thread density bound. Recently, the authors in~\cite{Headrick:2020gyq} investigated this issue in great detail. They proposed and proved several theorems on the existence of locking thread configurations in various situations, which will play a fundamental role in our study. For convenience, we collect the existence theorems of locking thread configurations that are necessary for our work from~\cite{Headrick:2020gyq} as follows. For more existence theorems and detailed technical proofs of the theorems, see the original paper~\cite{Headrick:2020gyq}.

Consider a $d$-dimensional compact Riemannian manifold-with-boundary $M$, for example, it can be a time slice of AdS$_{d + 1}$ spacetime, and then divide its boundary system $\partial M$ into adjacent non-overlapping subregions ${A_1}, \ldots ,{A_n}$, which are referred to as $elementary~regions$, satisfying ${A_i} \cap {A_j} = \emptyset $, $\mathop  \cup \limits_{i = 1}^n {A_i} = \partial M$. Accordingly, a $composite~ region$ is defined as the union of some certain elementary regions. It was shown in~\cite{ Headrick:2020gyq} that when we adopt the most traditional definition of the density of bit threads~\cite{Cui:2018dyq}, i.e., defining the thread density ρ as the total length of threads contained in a small ball divided by its volume: 
\be {\rho _v }\left( V \right) = \sum\limits_{i < j} {\left| {{{\vec v}_{ij}}} \right|}, \ee
 then we have the following locking theorems(following~\cite{ Headrick:2020gyq}, the multiflow describing the thread configuration under this definition of thread density is called a ${\nu _v}$ multiflow): 

{\bf Theorem 1.} There exists a ${\nu _v}$ multiflow that locks all the elementary regions ${A_i}$.~\footnote{Strictly speaking, this theorem was first proposed and proved in~\cite{Cui:2018dyq}, and restated in~\cite{ Headrick:2020gyq} in this systematic manner.}

{\bf Theorem 2.} There exists a ${\nu _v}$ multiflow that can lock all the elementary regions and all non-crossing composite regions simultaneously.

Here a $composite~ region$ is defined as the union of some certain elementary regions, and we are following the terminology from network theory: two boundary regions are said to cross if they partially overlap and do not cover the whole boundary. For example, $AB$ crosses $BC$, but does not cross $A$, $ABC$, or $D$. More explicitly, two regions $X$ and $Y$ do not cross if and only if at least one of the following conditions holds:
\be X \cap Y = \emptyset ,\quad X \subseteq Y,\quad Y \subseteq X,\quad X \cup Y = \partial M .\ee

It was further shown in~\cite{Headrick:2020gyq} that if we choose another reasonable definition of the density of bit threads, i.e., defining the thread density as the number per unit area intersecting a small disk, maximized over the orientation of the disk:
\be {\rho _a}\left( V \right) = \mathop {\max }\limits_{\hat n} \sum\limits_{i \prec j} {\left| {\hat n \cdot {{\vec v}_{ij}}} \right|}, \ee
where ${\hat n}$ is the unit vector normal to the small disk, then for the corresponding so-called ${\nu _a}$ multiflow, we have a more powerful locking theorem as follows:

{\bf Theorem 3.} There exists a ${\nu _a}$ multiflow that can lock two nested sequence sets simultaneously. 

Here the nested sequence set ${I_{{\rm{nest}}}}$ means that the elements in the set can be arranged in such an order that the former element is always contained by the next one, such as ${I_{{\rm{nest}}}} = \left\{ {A,AB,ABCD, \ldots } \right\}$.

%%%%%%%%%%%%%%%%%%%%%%%%%%%%%%%%%%%%%%%%%%%%%%%%%%%%%%%%%%%%%%%%%%%%%%

%%%%%%%%%%%%%%%%%%%%%%%%%%%%%%%%%%%%%%%%%%%%%%%%%%%%%%%%%%%%%%%%%%%%%%
\section{PEE proposal and locking bit thread configuration }\label{sec3}
%%%%%%%%%%%%%%%%%%%%%%%%%%%%%%%%%%%%%%%%%%%%%%%%%%%%%%%%%%%%%%%%%%%%%%
%%%%%%%%%%%%%%%%%%%%%%%%%%%%%%%%%%%%%%%%%%%%%%%%%%%%%%%%%%%%%%%%%%%%%%
%%%%%%%%%%%%%%%%%%%%%%%%%%%%%%%%%%%%%%%%%%%%%%%%%%%%%%%%%%%%%%%%%%%%%%

%%%%%%%%%%%%%%%%%%%%%%%%%%%%%%%%%%%%%%%%%%%%%%%%%%%%%%%%%%%%%%%%%%%%%%
\subsection{PEE (partial entanglement entropy) as CFF (component flow flux) }\label{subsec3.1}
%%%%%%%%%%%%%%%%%%%%%%%%%%%%%%%%%%%%%%%%%%%%%%%%%%%%%%%%%%%%%%%%%%%%%%
%%%%%%%%%%%%%%%%%%%%%%%%%%%%%%%%%%%%%%%%%%%%%%%%%%%%%%%%%%%%%%%%%%%%%%
%%%%%%%%%%%%%%%%%%%%%%%%%%%%%%%%%%%%%%%%%%%%%%%%%%%%%%%%%%%%%%%%%%%%%%

%%The motivation of our research arises from the recent study on the locking bit thread configurations under the different precise definitions of thread density bound and involving various types of sets of the boundary subregions~\cite{Headrick:2020gyq}. The terminology ``lock'' is borrowed from the network flow theory. A thread configuration that maximizes the thread flux through a region $A$ is said to lock $A$. Using the basic properties of bit threads, it is obvious to find a thread configuration that can lock a specified boundary subregion, and there is typically an infinite number of choices. However, one can further ask, can we find a locking thread configuration that can lock a specified set of subregions , i.e., maximize the thread fluxes through all these subregions simultaneously? The question becomes very nontrivial. Broadly speaking, it depends not only on the relative space position relations between these specified subregions, but also on the properties we assign to the bit threads, in particular, the precise definition of the thread density bound. %%

\begin{figure}[htbp]     \begin{center}
		\includegraphics[height=7.5cm,clip]{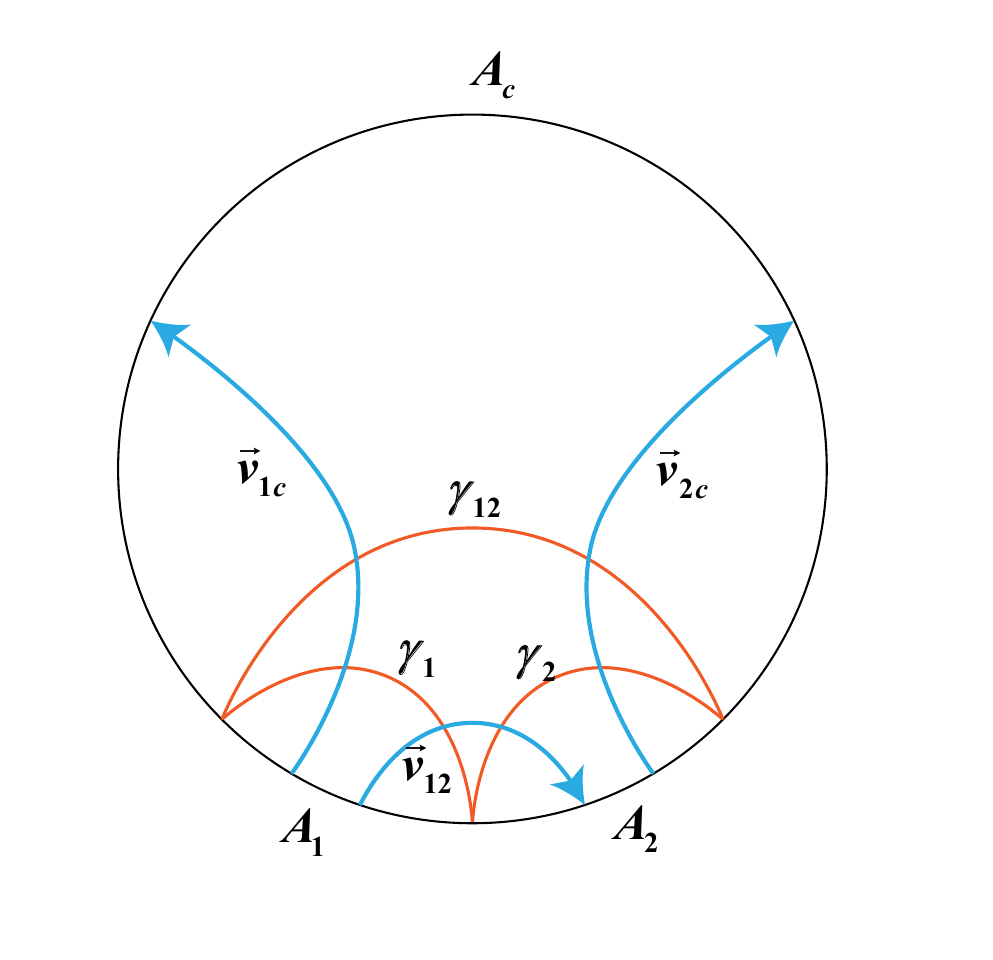}
		\caption{A locking bit thread configuration involving three elementary regions ${A_1}$, ${A_2}$, and ${A_c}$. It is described by a ${\nu _v}$ multiflow $V = \left\{ {{{\vec v}_{ij}}} \right\}$, in which each component flow ${\vec v_{ij}}$ is represented by a blue line. The red lines denote the RT surfaces associated with the involving boundary regions.}
		\label{fig1}
	\end{center}	
\end{figure}

In order to illustrate our statement, let us directly take the simplest case as an example. Considering a pure state quantum system described by a conformal field theory that has a holographic dual, we divide a subregion $A$ into two parts, ${{A_1}}$ and ${{A_2}}$, see figure \ref{fig1}. From the definition of the PEE, we know that ${s_A}\left( {{A_1}} \right)$ represents the contribution of ${A_1}$ to the entanglement entropy $S\left( A \right)$ between $A$ and its complement. Similarly, ${s_A}\left( {{A_2}} \right)$ represents the contribution of ${A_2}$ to $S\left( A \right)$. And we have
\be\label{pee}{s_A}\left( {{A_1}} \right) + {s_A}\left( {{A_2}} \right) = S\left( A \right) .\ee
As pointed out in~\cite{Kudler-Flam:2019oru}, this idea is very similar to the idea of bit threads in the physical picture. Let us investigate this situation more clearly from the perspective of the locking thread configuration.

Suppose we now want to construct a bit thread configuration that can lock a set of boundary subregions $I = \left\{ {{A_1},{A_2},{A_1}{A_2}} \right\}$ simultaneously (where we denote $A = {A_1} \cup {A_2} = {A_1}{A_2}$), described by a multiflow mathematically. Given the well-known zero-divergence property of multiflow, this is equivalent to requiring that a thread configuration maximizes the flux on the RT surfaces associated with each boundary subregion. Note that since the RT surface itself is an extremal surface with minimal area, according to the max flow-min cut theorem, this is also equivalent to require the thread configuration to satisfy that the thread flux on each involving RT surface is exactly equal to its area. From the perspective of RT surfaces, it is easy to see from the figure \ref{fig1} that the above locking problem is also equivalent to  requiring the bit thread configuration to lock the set $I = \left\{ {{A_1},{A_2},{A_c}} \right\}$(where we denote the complement of $A$ as ${A_c}$), this is because in a pure state, we have $S\left( A \right) = S\left( {{A_c}} \right)$. 

It turns out that, when we adopt the most traditional definition of the density of bit threads~\cite{Cui:2018dyq}, i.e., defining the thread density ρ as the total length of threads contained in a small ball divided by its volume, such a locking thread configuration always exists. %The multiflow describing the thread configuration under this definition of thread density is called a ${\nu _v}$ multiflow. Dividing the whole boundary system $\partial M$ into adjacent non-overlapping subregions ${A_1}, \cdots ,{A_n}$, which are referred to as $elementary~regions$ and satisfy ${A_i} \cap {A_j} = \emptyset $, $\mathop  \cup \limits_{i = 1}^n {A_i} = \partial M$,%
As reviewed in section~\ref{subsec2.2}, we have the following locking theorem~\cite{Cui:2018dyq,Headrick:2020gyq}:

{\bf Theorem 1.} There exists a ${\nu _v}$ multiflow that locks all the elementary regions ${A_i}$.

Therefore, by applying ${\bf Theorem~1}$, we can immediately assign a locking thread configuration as shown in figure \ref{fig1} for $I = \left\{ {{A_1},{A_2},{A_c}} \right\}$. One can see that, in this locking thread configuration, there are three independent $thread~bundles$ in total. Each thread bundle connects two distinct elementary regions ${A_i}$ and ${A_j}$, and is described by a $component~flow$ ${\vec v_{ij}}$ in the multiflow $V = \left\{ {{{\vec v}_{ij}}} \right\}$. From the properties of multiflow,
\be\label{gauss}\nabla  \cdot {\vec v_{ij}} = 0,\ee
\be\label{bypass}{\hat n_{{A_k}}} \cdot {\vec v_{ij}} = 0\,\,\,\,\,({\rm{for}}\,\,k \ne i,j{\rm{)}},\ee
we have
\be\label{f1c} F{\left( {{A_1}} \right)_{1c}} = F{\left( {{\gamma _1}} \right)_{1c}} = F{\left( {{\gamma _{12}}} \right)_{1c}} = F{\left( {{A_c}} \right)_{1c}},\ee
\be\label{f2c} F{\left( {{A_2}} \right)_{2c}} = F{\left( {{\gamma _2}} \right)_{2c}} = F{\left( {{\gamma _{12}}} \right)_{2c}} = F{\left( {{A_c}} \right)_{2c}},\ee
where we denote the RT surfaces associated with ${A_1}$, ${A_2}$ and ${A_{12}}$ as ${\gamma_1}$, ${\gamma_2}$ and ${\gamma_{12}}$ respectively, and $F{(\alpha )_{ij}} = \left| {\int_\alpha  {{{\vec v}_{ij}}} } \right| = \left| {\int_\alpha  {\sqrt h {{\hat n}_\alpha } \cdot {{\vec v}_{ij}}} } \right|$ represents the value of the flux of the bit threads described by the component flow ${\vec v_{ij}}$ passing through the $\alpha $ surface. $h$ is the determinant of the induced metric on the surface $\alpha $, and ${\hat n_\alpha }$ is the unit normal vector on surface $\alpha $. One can see that due to the divergenceless constraint~(\ref{gauss}) the flux of the thread bundle ${\vec v_{ij}}$ does not change when the threads start from the elementary region ${A_i}$, pass through each intersecting RT surface, and come back to another elementary region ${A_j}$.

Denoting $N\left( \alpha  \right) = Flu{x_{{\rm{locking}}}}\left( \alpha  \right)$ as the total thread flux through the surface $\alpha $ of interest in this locking thread configuration, and $S\left( \alpha  \right)$ as the entanglement entropy of $\alpha $ (in the spirit of the surface/state correspondence~\cite{Miyaji:2015yva,Miyaji:2015fia}), then considering the bit threads on the RT surface ${\gamma_{12}}$ (which also corresponds to ${A_c}$), we have
\be F{\left( {{\gamma _{12}}} \right)_{1c}} + F{\left( {{\gamma _{12}}} \right)_{2c}} = N\left( {{\gamma _{12}}} \right) \equiv S\left( {{\gamma _{12}}} \right) ,\ee
or, by~(\ref{f1c})~(\ref{f2c}), and $N\left( {{\gamma _{12}}} \right) = N\left( A \right) \equiv S\left( A \right)$, equivalently,
\be F{\left( {{A_1}} \right)_{1c}} + F{\left( {{A_2}} \right)_{2c}} = S\left( A \right) ,\ee
as expected. Unsurprisingly, we see that this is exactly accordant with the definition of the PEE~(\ref{pee}), as long as we interpret the PEE ${s_A}\left( {{A_1}} \right)$ as the component flow flux $F{\left( {{A_i}} \right)_{ic}}$ in the locking thread configuration, i.e.,
\be\label{int}{s_A}\left( {{A_i}} \right) = {F^L}{\left( {{A_i}} \right)_{ic}} .\ee
(where the superscript $L$ indicates the locking thread configuration, we will omit it in the rest of this paper for convenience). If we rewrite the PEE ${s_A}\left( {{A_1}} \right)$ as the form with permutation symmetry ${s_A}\left( {{A_i}} \right) = P\left( {{A_i},{A_c}} \right) \equiv {P_{ic}}$ following~(\ref{per}), and simply denote $F{(\alpha )_{ij}} = \left| {\int_\alpha  {{{\vec v}_{ij}}} } \right| = \left| {\int_\alpha  {\sqrt h {{\hat n}_\alpha } \cdot {{\vec v}_{ij}}} } \right|$ as ${F_{ij}}$, since the flux of each component flow ${\vec v_{ij}}$ is independent of the surfaces it passes through, our claim can also be express as
\be\label{int2}{P_{ic}} = {F_{ic}} .\ee
Recall that the motivation for rewriting the PEE ${s_A}\left( {{A_i}} \right)$ as ${P_{ic}}$ in~(\ref{per}) is that physically it  represents the correlation between the subset ${A_i}$ and ${A_c}$. The identification~(\ref{int2}) indicates that the picture of locking bit thread configuration can intuitively describe this idea. Furthermore, as we will see later, this natural interpretation will provide a nice proof for the nontrivial PEE proposal in the holographic framework.

%%%%%%%%%%%%%%%%%%%%%%%%%%%%%%%%%%%%%%%%%%%%%%%%%%%%%%%%%%%%%%%%%%%%%%

\subsection{The bit thread interpretation of PEE proposal}\label{subsec3.2}
%%%%%%%%%%%%%%%%%%%%%%%%%%%%%%%%%%%%%%%%%%%%%%%%%%%%%%%%%%%%%%%%%%%%%%
Let us start with a brief review of the PEE proposal. The author in~\cite{Wen:2019ubu} proposed that, dividing a subsystem $A$ in a quantum system into three parts (for convenience we denote them as ${A_1}$, ${A_2}$, and ${A_3}$ in order), the PEE ${s_A}\left( {{A_2}} \right)$ in the middle part can be obtained by the following proposal:
\be\label{pro}{s_A}\left( {{A_2}} \right) = {\textstyle{1 \over 2}}\left( {{S_{12}} + {S_{23}} - {S_1} - {S_3}} \right) ,\ee
where and hereafter we use the shorthand, such as ${S_1} = S\left( {{A_1}} \right)$, ${S_{12}} = S\left( {{A_1}{A_2}} \right){\rm{ = }}S\left( {{A_1} \cup {A_2}} \right)$, etc. Note that since we can freely choose the position of ${A_2}$ in $A$, in principle, we can calculate the contribution of any part of $A$ to the entanglement entropy $S\left( A \right)$, in terms of the linear combination of the entropies of several subregions.

\begin{figure}[htbp]     \begin{center}
		\includegraphics[height=5.5cm,clip]{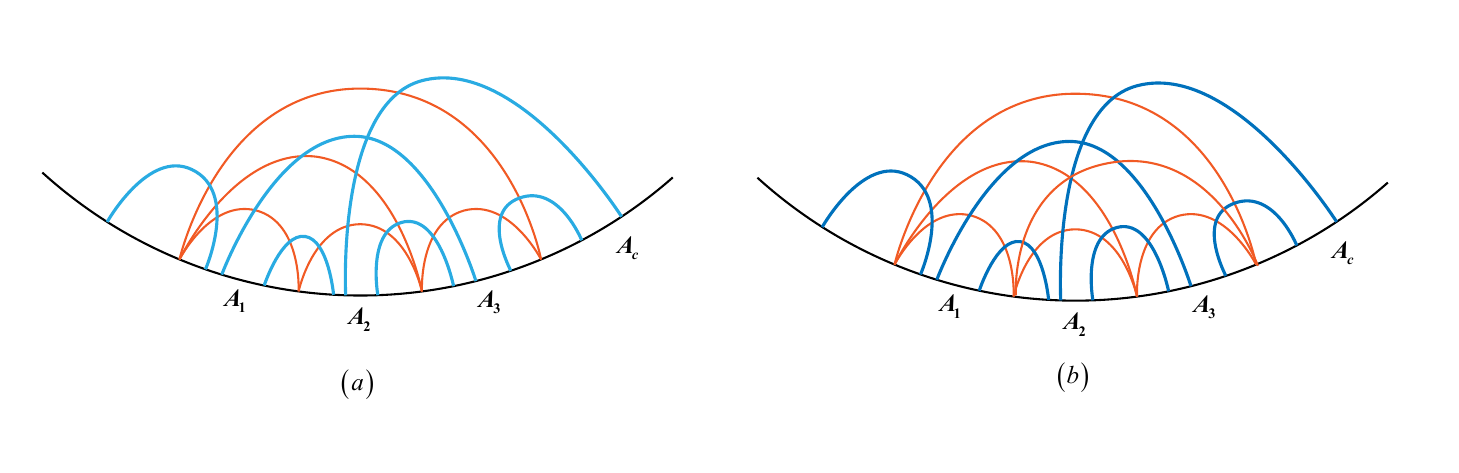}
		\caption{Two kinds of multiflows in realizing the locking thread configuration characterizing the entanglement between a tripartite subregion $A$ and its complement ${A_c}$: (a) in the scheme by Theorem 2, the locking thread configuration is described by a ${\nu _v}$ multiflow, which can only lock five RT surfaces; (b) in the scheme by Theorem 3, the locking thread configuration is described by a ${\nu _a}$ multiflow, which can lock six RT surfaces. We draw the latter multiflow in deep blue to distinguish them.}
		\label{fig2}
	\end{center}	
\end{figure}

Now, we will investigate this tripartite subsystem from the viewpoint that the PEE can be interpreted as the component flow flux in the locking bit thread configuration, and show that it can derive the PEE proposal~(\ref{pro}). Similar to the previous section, we first need to guarantee the existence of the locking thread configuration required. Naively, one may want to use the same definition of thread density as in the previous and still consider the locking thread configuration constructed by a ${\nu _v}$ multiflow. Indeed, as reviewed in section~\ref{subsec2.2}, it has been proved that there exists a more powerful locking theorem for the ${\nu _v}$ multiflow as follows~\cite{Headrick:2020gyq}:

{\bf Theorem 2.} There exists a multiflow that can lock all the elementary regions and all non-crossing composite regions simultaneously.

%%Here a $composite~ region$ is defined as the union of some certain elementary regions, and we are following the terminology from network theory: two boundary regions are said to cross if they partially overlap and do not cover the whole boundary. For example, $AB$ crosses $BC$, but does not cross $A$, $ABC$, or $D$. More explicitly, two regions $X$ and $Y$ do not cross if and only if at least one of the following conditions holds:\be X \cap Y = \emptyset ,\quad X \subseteq Y,\quad Y \subseteq X,\quad X \cup Y = \partial M\ee
%%%

Therefore, as long as we select a set of specified regions satisfying the non-crossing condition, we can always find a thread configuration that can lock all these specified regions simultaneously. As shown in the figure \ref{fig2}(a)~\footnote{Such a locking thread configuration has been explicitly constructed in~\cite{Lin:2020yzf}. }, we choose such a non-crossing set as
\be I = \left\{ {{A_1},{A_2},{A_3},{A_1}{A_2},{A_1}{A_2}{A_3}} \right\} .\ee
This can also be seen intuitively from the RT surfaces associated with the specified regions, because if the two specified regions do not cross, their corresponding RT surfaces will not intersect, rather, they are separated from each other, or one of them surrounds the other one. Note that based on the properties of bit threads, this locking thread configuration can also lock the complements of each element in $I$, such as ${A_c}$, the complement of ${A_1}{A_2}{A_3}$. Tracing the source of the bit threads on each involving RT surface in this locking thread configuration, we obtain
\be\label{con1}\begin{array}{l}
	{F_{1c}} + {F_{12}} + {F_{13}} = {S_1}\\
	{F_{1c}} + {F_{13}} + {F_{2c}} + {F_{23}} = {S_{12}}\\
	{F_{1c}} + {F_{2c}} + {F_{3c}} = {S_{123}}\\
	{F_{12}} + {F_{23}} + {F_{2c}} = {S_2}\\
	{F_{13}} + {F_{23}} + {F_{3c}} = {S_3}
\end{array} .\ee
Again, due to the properties of multiflow~(\ref{gauss}) and (\ref{bypass}), the flux of thread bundle ${\vec v_{ij}}$ does not change when the threads start from the elementary region ${A_i}$, pass through each intersecting RT surface, and come back to another elementary region ${A_j}$. Therefore we can simply denote $F{(\alpha )_{ij}} = \left| {\int_\alpha  {{{\vec v}_{ij}}} } \right| = \left| {\int_\alpha  {\sqrt h {{\hat n}_\alpha } \cdot {{\vec v}_{ij}}} } \right|$ as ${F_{ij}}$.

However, there is a subtle problem here. Due to the requirement of the non-crossing condition, there are only five non-intersecting RT surfaces that are locked in our configuration, while there are six component flows in a locking thread configuration describing this tripartite case. Our aim is to solve the component flow fluxes and see if they can exactly describe the PEE, however, in this case, we have six variables, but only five constraints. It seems that we get into trouble in further interpreting the PEE as the component flow flux, because if we want to add another constrain, we must introduce a new member that crosses the other specified regions. For example, based on the reasonable requirement of symmetry, one would like to add another RT surface ${\gamma _{23}}$ associated with ${A_2}{A_3}$ to be locked in the figure, however, it obviously intersects the surface ${\gamma _{12}}$, since ${A_2}{A_3}$ crosses ${A_1}{A_2}$. It should be emphasized that ${\bf Theorem~2}$ does not mean that one can never find a ${\nu _v}$ multiflow that can lock a crossing set. What it means is that one cannot always find a ${\nu _v}$ multiflow that can lock a specified crossing set. However, since the PEE proposal should be valid for any tripartition of $A$ in principle, if our bit thread interpretation of the PEE is correct, one should always be able to find the appropriate locking thread configuration enough to solve the value of component flow flux in every case.

Happily, there is a ready solution to this problem. It has been shown in~\cite{Headrick:2020gyq} that if we choose another reasonable definition of the density of bit threads, i.e., defining the thread density as the number per unit area intersecting a small disk, maximized over the orientation of the disk, then for the corresponding so-called ${\nu _a}$ multiflow, as reviewed in section~\ref{subsec2.2}, we have a more powerful locking theorem as follows (for details see the original paper~\cite{Headrick:2020gyq}):

{\bf Theorem 3.} There exists a ${\nu _a}$ multiflow that can lock two nested sequence sets simultaneously. 

%The nested sequence set ${I_{{\rm{nest}}}}$ means that the elements in the set can be arranged in such an order that the former element is always contained by the next one, such as ${I_{{\rm{nest}}}} = \left\{ {A,AB,ABCD, \ldots } \right\}$.%

Now we can select two nested sequence sets as $\left\{ {{A_1},{A_1}{A_2},{A_1}{A_2}{A_c}} \right\}$ and $\left\{ {{A_2},{A_2}{A_3},{A_1}{A_2}{A_3}} \right\}$. Then according to ${\bf Theorem~3}$, we can always find a ${\nu _a}$ multiflow that can lock a specified set of subregions $I = \left\{ {{A_1},{A_1}{A_2},{A_1}{A_2}{A_c}} \right\} \cup \left\{ {{A_2},{A_2}{A_3},{A_1}{A_2}{A_3}} \right\}$. The key point is that the complement of ${A_1}{A_2}{A_c}$ is exactly ${A_3}$. In other words, the RT surface corresponding to ${A_1}{A_2}{A_c}$ is exactly ${\gamma _3}$. Therefore, in this way, now we can find a locking thread configuration as shown in figure \ref{fig2}(b), which locks one more RT surface that surrounds ${A_2}$ and ${A_3}$, accordingly, we can finally introduce a new constraint:
\be\label{con2} {F_{3c}} + {F_{13}} + {F_{2c}} + {F_{12}} = {S_{23}} .\ee
In order to analyze the structure of the solution, the above equations~(\ref{con1}) and (\ref{con2}) can be written in the form of a matrix equation, i.e.,
\be\left( {\begin{array}{*{20}{c}}
		1&1&1&0&0&0\\
		1&0&1&1&1&0\\
		1&0&0&1&0&1\\
		0&1&0&1&1&0\\
		0&0&1&0&1&1\\
		0&1&1&1&0&1
\end{array}} \right)\left[ {\begin{array}{*{20}{c}}
		{{F_{1c}}}\\
		{{F_{12}}}\\
		{{F_{13}}}\\
		{{F_{2c}}}\\
		{{F_{23}}}\\
		{{F_{3c}}}
\end{array}} \right] = \left[ {\begin{array}{*{20}{c}}
		{{S_1}}\\
		{{S_{12}}}\\
		{{S_{123}}}\\
		{{S_2}}\\
		{{S_3}}\\
		{{S_{23}}}
\end{array}} \right] .\ee
It is easy to verify that the determinant of the matrix is not zero and the matrix has full rank, therefore, the solution of the equations exists and is unique. We immediately obtain the solution as
\be\left[ {\begin{array}{*{20}{c}}
		{{F_{1c}}}\\
		{{F_{12}}}\\
		{{F_{13}}}\\
		{{F_{2c}}}\\
		{{F_{23}}}\\
		{{F_{3c}}}
\end{array}} \right] = \left( {\begin{array}{*{20}{c}}
		{\frac{1}{2}}&0&{\frac{1}{2}}&0&0&{ - \frac{1}{2}}\\
		{\frac{1}{2}}&{ - \frac{1}{2}}&0&{\frac{1}{2}}&0&0\\
		0&{\frac{1}{2}}&{ - \frac{1}{2}}&{ - \frac{1}{2}}&0&{\frac{1}{2}}\\
		{ - \frac{1}{2}}&{\frac{1}{2}}&0&0&{ - \frac{1}{2}}&{\frac{1}{2}}\\
		0&0&0&{\frac{1}{2}}&{\frac{1}{2}}&{ - \frac{1}{2}}\\
		0&{ - \frac{1}{2}}&{\frac{1}{2}}&0&{\frac{1}{2}}&0
\end{array}} \right)\left[ {\begin{array}{*{20}{c}}
		{{S_1}}\\
		{{S_{12}}}\\
		{{S_{123}}}\\
		{{S_2}}\\
		{{S_3}}\\
		{{S_{23}}}
\end{array}} \right] ,\ee
or
\be\label{sys1}\left[ {\begin{array}{*{20}{c}}
		{{F_{1c}}}\\
		{{F_{12}}}\\
		{{F_{13}}}\\
		{{F_{2c}}}\\
		{{F_{23}}}\\
		{{F_{3c}}}
\end{array}} \right] = \left[ {\begin{array}{*{20}{c}}
		{{\textstyle{1 \over 2}}\left( {{S_1} + {S_{123}} - {S_{23}}} \right)}\\
		{{\textstyle{1 \over 2}}\left( {{S_1} + {S_2} - {S_{12}}} \right)}\\
		{{\textstyle{1 \over 2}}\left( {{S_{12}} + {S_{23}} - {S_2} - {S_{123}}} \right)}\\
		{{\textstyle{1 \over 2}}\left( {{S_{12}} + {S_{23}} - {S_1} - {S_3}} \right)}\\
		{{\textstyle{1 \over 2}}\left( {{S_2} + {S_3} - {S_{23}}} \right)}\\
		{{\textstyle{1 \over 2}}\left( {{S_3} + {S_{123}} - {S_{12}}} \right)}
\end{array}} \right] .\ee
In particular, we have
\be\label{sol}{F_{2c}} = {\textstyle{1 \over 2}}\left( {{S_{12}} + {S_{23}} - {S_1} - {S_3}} \right) ,\ee
which is exactly the same as the PEE proposal~(\ref{pro}) put forward in~\cite{Wen:2019ubu}, and consistent with our interpretation in~(\ref{int}), i.e.,
\be{s_A}\left( {{A_2}} \right) = {F_{2c}} .\ee
Furthermore, it can be checked that the other formulas are also in complete agreement with the formulas about the PEE in~\cite{Wen:2019ubu}, given our identification~(\ref{int}). In a word, from the perspective of locking bit thread configuration, the whole picture of the PEE obtains an intuitive and natural interpretation.

A few comments: one may worry that we change the definition of thread density in this section, however, as point out in~\cite{Headrick:2020gyq}, changing the density definition of bit threads from the ${\nu _v}$ version to the ${\nu _a}$ version does not affect the really essential characteristics of bit threads, i.e., as long as we still require the thread density to be less than $1$ everywhere in the bulk, the maximum flux of threads through any surface is still equal to the area of the minimal bulk surface homologous to it, thus we can still reproduce the desired RT formula.

%%%%%%%%%%%%%%%%%%%%%%%%%%%%%%%%%%%%%%%%%%%%%%%%%%%%%%%%%%%%%%%%%%%%%%
\subsection{Balanced PEE and EoP}\label{subsec3.3}
%%%%%%%%%%%%%%%%%%%%%%%%%%%%%%%%%%%%%%%%%%%%%%%%%%%%%%%%%%%%%%%%%%%%%%
\begin{figure}[htbp]     \begin{center}
		\includegraphics[height=9cm,clip]{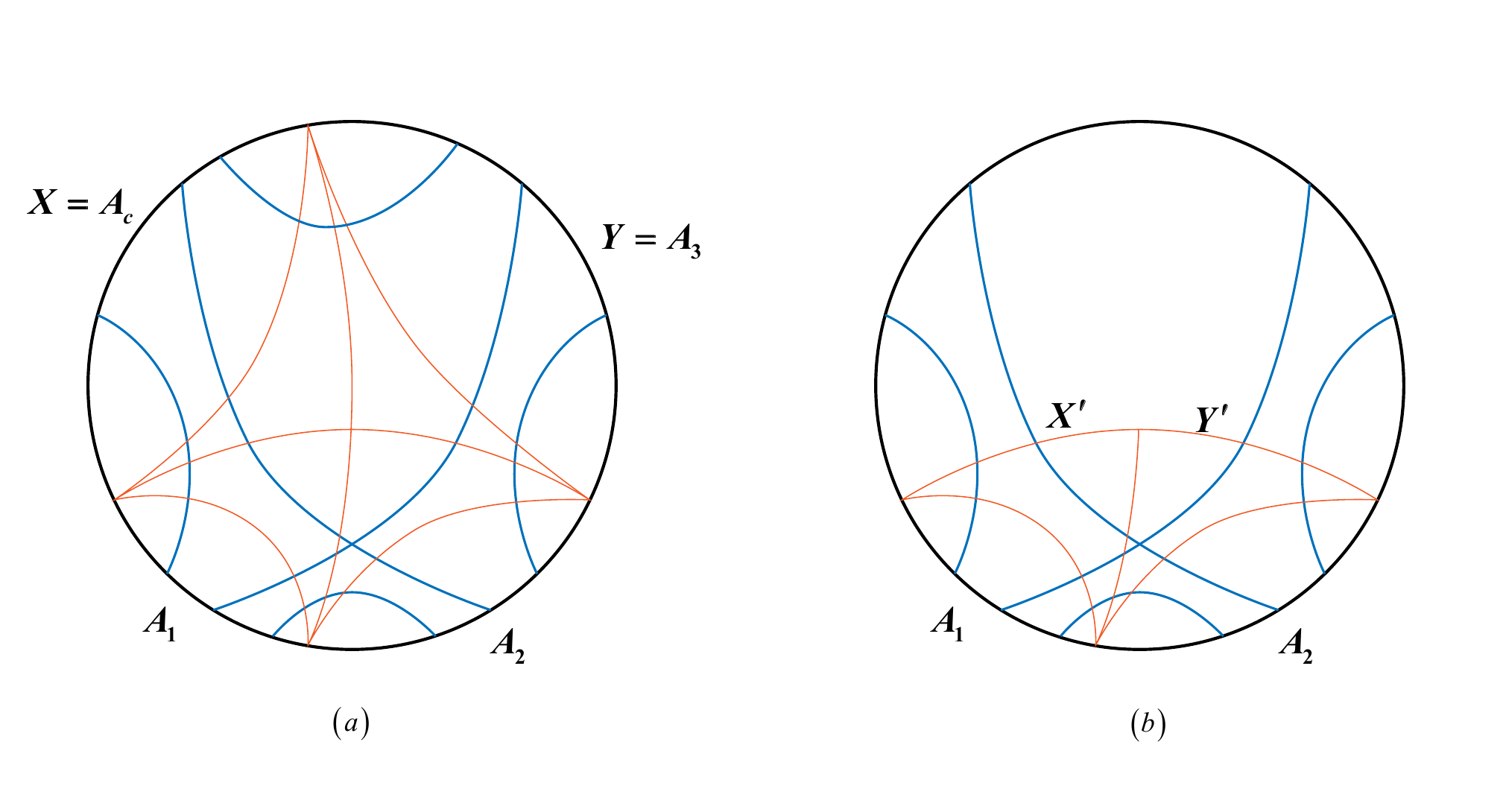}
		\caption{
			(a)	The locking bit thread configuration corresponding to the BPE, in which ${F_{13}} = {F_{2c}}$.
			(b)	The locking bit thread configuration corresponding to the holographic EoP, in which there are only $5$ component flows.
		}
		\label{fig3}
	\end{center}	
\end{figure}

Recently, a new quantity, called balanced partial entanglement (BPE) was proposed in~\cite{Wen:2021qgx} to measure the correlation between two parts ${A_1}$ and ${A_2}$ of a bipartite system ${A_1}{A_2}$, and is found to be equal to the area of the entanglement wedge cross section (EWCS) of this bipartite system ${A_1}{A_2}$ in the holographic context. The entanglement wedge cross section ${\Gamma _{{A_1}:{A_2}}}$ of a bipartite system ${A_1}{A_2}$ is a surface of minimal area anchored to the boundary of the entanglement wedge ${W_{{A_1}{A_2}}}$ of ${A_1} {A_2}$ (i.e., the bulk region surrounded by ${A_1}{A_2}$ and its corresponding RT surface $\gamma \left( {{A_1}{A_2}} \right)$), such that ${\Gamma _{{A_1}:{A_2}}}$ partitions ${W_{{A_1}{A_2}}}$ into a region that is entirely adjacent to ${A_1}$ and another region that is entirely adjacent to ${A_2}$. 

The definition of the BPE can be expressed as follows: first finding two auxiliary systems (denoted as $X$ and $Y$) for the bipartite system ${A_1}{A_2}$ such that the whole ${A_1}{A_2}YX$ is in a pure state, and further requiring that
\be\label{bpe}{s_{{A_1}X}}\left( {{A_1}} \right) = {s_{{A_2}Y}}\left( {{A_2}} \right) .\ee
From the basic properties of the PEE, it is easy to obtain that this is also equivalent to requiring that
\be\label{bpe2}{P_{{A_1}Y}} = {P_{{A_2}X}} ,\ee
then the BPE $ BPE\left( {{A_1},{A_2}} \right)$ is defined as the ${s_{{A_1}X}}\left( {{A_1}} \right)$ satisfying the above condition~(\ref{bpe}) (in general taking its minimal possible value), i.e.,
\be BPE\left( {{A_1},{A_2}} \right) = {\left. {{s_{{A_1}X}}\left( {{A_1}} \right)} \right|_{balance}} .\ee
It was found that~\cite{Wen:2021qgx}, in the holographic framework, since the whole boundary system is in a pure state, the above process is equivalent to selecting a special point $Q$ on the boundary to divide the complement of ${A_1}{A_2}$ into two regions $X$ and $Y$, such that the balance requirement~(\ref{bpe}) is satisfied, as shown in figure \ref{fig3}(a). And it turns out that the $BPE\left( {{A_1},{A_2}} \right)$ is exactly equal to the area of the entanglement wedge cross section of ${A_1}{A_2}$.

Interestingly, if we regard $Y$ as ${A_3}$, and $X$ as ${A_c}$, then the figure \ref{fig3}(a) is exactly the same as the figure \ref{fig2} we considered in the previous subsection, we can thus use the same locking thread configuration to explain its detailed entanglement structure. Moreover, from the perspective of  our interpretation~(\ref{int2}), the above balance process~(\ref{bpe2}) is essentially selecting a special ${A_3}$ region in the boundary system, such that the locking thread configuration in the previous subsection happens to satisfy: 
\be{F_{13}} = {F_{2c}} .\ee.

On the other hand, recently in~\cite{Lin:2020yzf}, the locking thread configuration is also be utilized to provide an interpretation for the so-called entanglement of purification (EoP)~\cite{EOP}, which is also equal to the area of the entanglement wedge cross section in the holographic context~\cite{Takayanagi:2017knl,Nguyen:2017yqw}, as shown in figure \ref{fig3}(b). 

Since both the PEE and the EoP are supposed to be dual to the same geometric quantity~\footnote{There are also other quantum information theoretical quantities proposed to be associated with the entanglement wedge cross section, such as the reflected entropy~\cite{Dutta:2019gen}, the logarithmic negativity~\cite{Kudler-Flam:2018qjo,Kusuki:2019zsp}, the ``odd entropy''~\cite{Tamaoka:2018ned}, the ``differential purification''~\cite{Espindola:2018ozt}, etc. There are also other papers investigating the holographic entanglement of purification from the view point of bit thread, such as~\cite{Du:2019emy,Bao:2019wcf,Harper:2019lff,Agon:2018lwq,Hubeny:2018bri}.}, it is natural to ask whether these two viewpoints can be related, or whether there exists some contradiction between them. In the following, we will utilize the viewpoint of locking bit thread configuration to show a harmonious picture for these two coincidences.

\begin{figure}[htbp]     \begin{center}
		\includegraphics[height=6.3cm,clip]{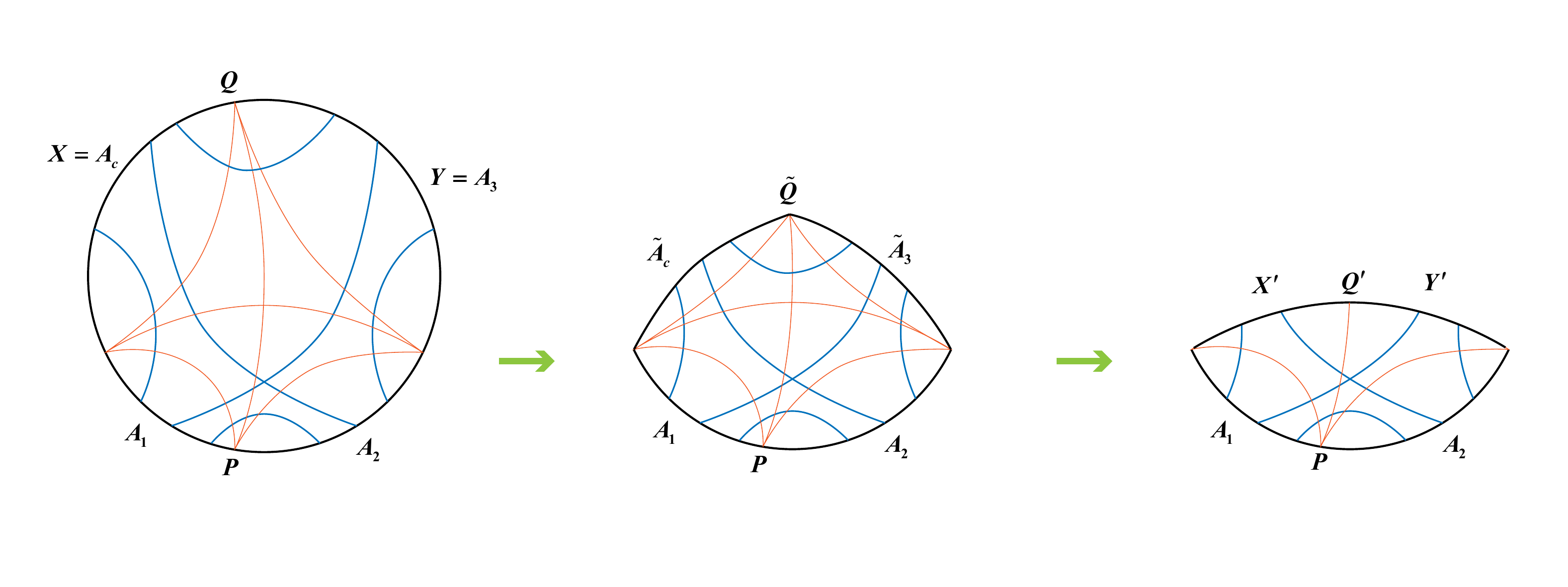}
		\caption{Imagine we gradually move the dividing point $Q$ of $XY$ (or ${A_c}{A_3}$) to the dividing point $Q'$ of $X'Y'$ (or ${A'_c}{A'_3}$) along the geodesic $PQ$ and accordingly push $XY$ into the bulk until it reaches $X'Y'$, while keeping the locking behavior of the thread configuration all the time. }
		\label{fig4}
	\end{center}	
\end{figure}

The key point is shown in the figure \ref{fig4}, imagining we gradually move the dividing point $Q$ of $XY$ (or ${A_c}{A_3}$) to the dividing point $Q'$ of $X'Y'$ (or ${A'_c}{A'_3}$) and accordingly push $XY$ into the bulk until it reaches $X'Y'$, while keeping the locking behavior of the thread configuration all the time. At the beginning, $Q$ is a special point specifying a special partner ${A_3}$ for ${A_1}$ and ${A_2}$ such that the locking thread configuration for $I = \left\{ {{A_1},{A_1}{A_2},{A_1}{A_2}{A_c}} \right\} \cup \left\{ {{A_2},{A_2}{A_3},{A_1}{A_2}{A_3}} \right\}$ satisfies ${F_{13}} = {F_{2c}}$. Next, we move $Q$ along the geodesic connecting $Q$ and the dividing point $P$ of ${A_1}$ and ${A_2}$, which is also the RT surface associated with ${{A_2}{A_3}}$, then in the spirit of surface/state correspondence~\cite{Miyaji:2015yva,Miyaji:2015fia}, we can regard ${A_1}{A_2}{\tilde A_3}{\tilde A_c}$ as a new holographic manifold. The locking thread configuration will also change accordingly to lock the set $I = \left\{ {{A_1},{A_1}{A_2},{A_1}{A_2}{{\tilde A}_c}} \right\} \cup \left\{ {{A_2},{A_2}{{\tilde A}_3},{A_1}{A_2}{{\tilde A}_3}} \right\}$, in which in general we no longer have ${F_{13}} = {F_{2c}}$. Finally, when $Q$ reaches $Q'$ on the ${\gamma _{12}}$ surface, ${\tilde A_3}$ itself becomes a minimal extremal surface $Y'$, while ${\tilde A_c}$ becomes the minimal extremal surface $X'$, and we find that the thread bundle described by ${\vec v_{3c}}$ connecting ${\tilde A_3}$ and ${\tilde A_c}$ disappears, thus ${F_{3c}}$ becomes zero. Indeed, this is reasonable because in the idea of surface/state correspondence, there should be no internal entanglement within a minimal extremal surface, and now ${\tilde A_3}$ and ${\tilde A_c}$ are located in the same minimal extremal surface ${\gamma _{12}}$. It turns out that the system of equations~(\ref{con1})(\ref{con2}) has one less unknown, but correspondingly, since now the RT surface ${\gamma _{3c}}$ associated with ${\tilde A_3}{\tilde A_c}$ is no longer independent of ${\gamma _3}$ and ${\gamma _c}$, the number of the constraints is also reduced by one. Therefore, we see that everything is consistent. At the beginning, we have $BPE\left( {{A_1},{A_2}} \right) = {F_{13}} + {F_{12}}$ equal to the area of EWCS of ${A_1}{A_2}$, while at the end it is the $EoP\left( {{A_1},{A_2}} \right) = {F_{13}} + {F_{12}} + {F_{2c}}$ that equals the area of EWCS of ${A_1}{A_2}$.

One can explicitly solve the component flow fluxes at the final stage to see the change of the fluxes more clearly. This is tantamount to deleting one equation ${F_{1c}} + {F_{13}} + {F_{2c}} + {F_{23}} = {S_{12}}$ and one unknown ${F_{3c}}$ from the previous system, and we obtain
\be\left( {\begin{array}{*{20}{c}}
		1&1&1&0&0\\
		1&0&0&1&0\\
		0&1&0&1&1\\
		0&0&1&0&1\\
		0&1&1&1&0
\end{array}} \right)\left[ {\begin{array}{*{20}{c}}
		{{F_{1c}}}\\
		{{F_{12}}}\\
		{{F_{13}}}\\
		{{F_{2c}}}\\
		{{F_{23}}}
\end{array}} \right] = \left[ {\begin{array}{*{20}{c}}
		{{S_1}}\\
		{{S_{123}}}\\
		{{S_2}}\\
		{{S_3}}\\
		{{S_{23}}}
\end{array}} \right] ,\ee
and
\be\label{sys2}\left[ {\begin{array}{*{20}{c}}
		{{F_{1c}}}\\
		{{F_{12}}}\\
		{{F_{13}}}\\
		{{F_{2c}}}\\
		{{F_{23}}}
\end{array}} \right] = \left[ {\begin{array}{*{20}{c}}
		{\frac{1}{2}\left( {{S_1} + {S_{123}} - {S_{23}}} \right)}\\
		{\frac{1}{2}\left( {{S_1} - {S_{123}} + {S_2} - {S_3}} \right)}\\
		{\frac{1}{2}\left( { - {S_2} + {S_3} + {S_{23}}} \right)}\\
		{\frac{1}{2}\left( { - {S_1} + {S_{123}} + {S_{23}}} \right)}\\
		{\frac{1}{2}\left( {{S_2} + {S_3} - {S_{23}}} \right)}
\end{array}} \right] .\ee
Note that at this final stage, we have
\be{S_{123}} + {S_3} = {S_{12}} .\ee
Therefore, (\ref{sys2}) is consistent with (\ref{sys1}), that is, in this ``pushing'' process, the dependence of each component flow flux on the subregion entropies does not change. However, because the entropies of regions involving ${\tilde A_3}$ and ${\tilde A_c}$ are changing during this process, the component flow fluxes are also changing accordingly. In particular, ${F_{12}}$ is constant during this process, as one would expect in physics that the purification operation should not change the correlation between ${A_1}$ and ${A_2}$.

%%%%%%%%%%%%%%%%%%%%%%%%%%%%%%%%%%%%%%%%%%%%%%%%%%%%%%%%%%%%%%%%%%%%%%
\section{More general PEE proposal}\label{sec4}
%%%%%%%%%%%%%%%%%%%%%%%%%%%%%%%%%%%%%%%%%%%%%%%%%%%%%%%%%%%%%%%%%%%%%%

%%%%%%%%%%%%%%%%%%%%%%%%%%%%%%%%%%%%%%%%%%%%%%%%%%%%%%%%%%%%%%%%%%%%%%
\subsection{A locking scheme for deriving the generalized PEE proposal }\label{subsec4.1}
%%%%%%%%%%%%%%%%%%%%%%%%%%%%%%%%%%%%%%%%%%%%%%%%%%%%%%%%%%%%%%%%%%%%%%

By utilizing the language of bit threads, it is natural to further study the situation that the subsystem $A$ is divided into more than three parts. According to the idea that the component flow flux in a locking thread configuration is identified as the PEE, one can study the expression of the PEE in these generalized situations. It turns out that we can exactly reproduce the generalized PEE proposal in~\cite{Kudler-Flam:2019oru}.\footnote{However, we should forewarn that there will be some subtleties in this generalization. As shown in~\cite{Headrick:2020gyq}, it seems that bit threads are not able to lock an arbitrary set of specified subregions according to our current understanding of bit threads. However, we will continue to point out this interesting coincidence between the locking thread configuration and the PEE proposal, and put the issues this coincidence implies in later discussion.}

\begin{figure}[htbp]     \begin{center}
		\includegraphics[height=5.7cm,clip]{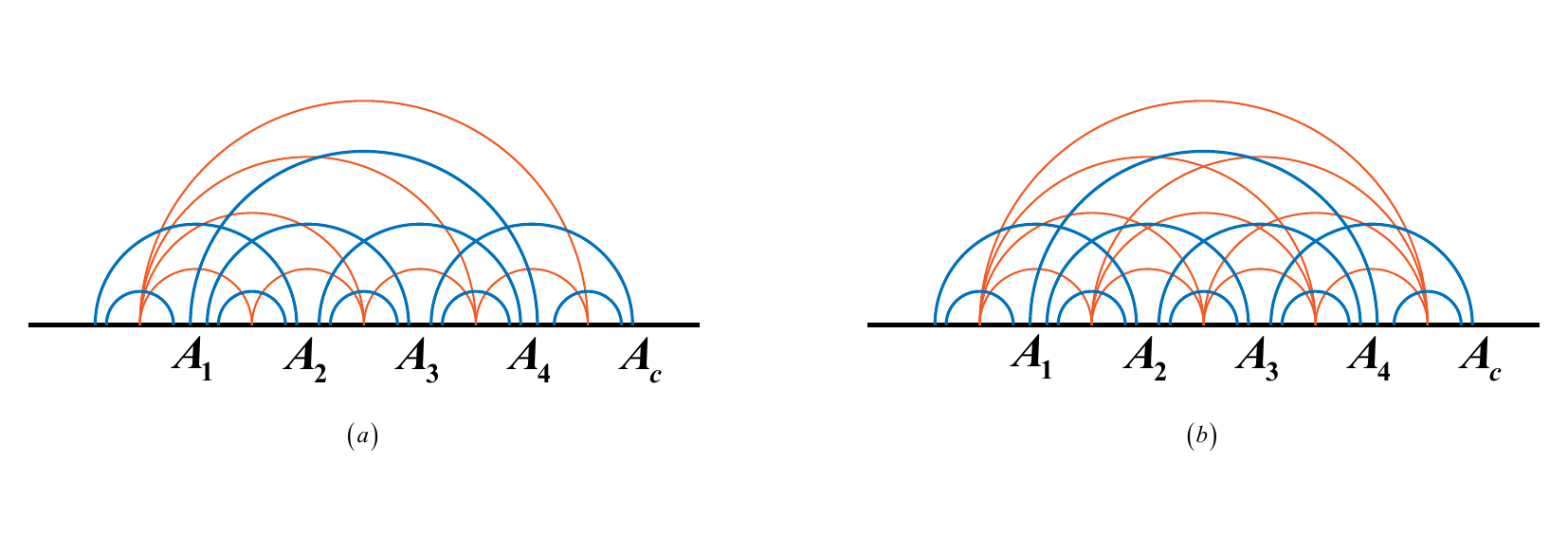}
		\caption{The locking thread configuration for the case that the subregion $A$ is divided into $4$ parts. There are $10$ independent component flows (denoted by blue lines) necessarily. (a) is the non-crossing type locking thread configuration, in which there are only $7$ RT surfaces (denoted by red lines) involved. In (b), we propose a natural scheme to add more RT surfaces so as to match the number of the component flows.}
		\label{fig5}
	\end{center}	
\end{figure}

Consider dividing the subregion $A$ into $n$ parts. For convenience, from now on, we denote the elementary region contained in region $A$ as ${A_i}$, where $i$ takes $1$ to $n$ in left-to-right (or anticlockwise) order as shown in figure \ref{fig5}, while we denote the rest elementary region as ${A_c}$ since it is the complement of $A$ region. As an example, we illustrate the case of $n=4$ in figure \ref{fig5}. In this case, the total number of independent thread bundles (which is characterized by the component flows) representing the entanglement between each pair of elementary regions is equal to the sum of an arithmetic sequence, i.e.,
\be n + \left( {n - 1} \right) +  \cdots  + 1 = \frac{{n\left( {n + 1} \right)}}{2} .\ee
In particular, there are totally $4{\rm{ + }}3{\rm{ + }}2{\rm{ + }}1{\rm{ = }}10$ component flows in the figure \ref{fig5}. Once again, we are faced with the same problem as in section~\ref{sec3}. The number of RT surfaces, which is corresponding to the constraints, contained in the non-crossing type locking thread configuration constructed by ${\bf Theorem~2}$, is less than the number of the independent thread bundles, as shown in the figure \ref{fig5}(a). And this number difference between the RT surfaces contained in the non-crossing type locking thread configuration and the involving independent thread bundles will continue to increase when $n$ becomes larger. Therefore, the first problem we face in the generalization is how to make the number of the RT surfaces (or constrains) involved in the locking thread configuration exactly match the number of the component flows (or unknowns), so as to exactly solve the value of each component flow flux. Noting that the number of the component flows happens to be the sum of a sequence with an arithmetic difference of $1$, a natural approach is to add the RT surfaces as shown in figure \ref{fig5}(b)~\footnote{To avoid misunderstanding, it is worth reiterating that what we really mean here by ``adding an RT surface'' is adding an additional requirement that the surface is locked by the thread configuration. Of course, the surface is always ``there'' regardless of whether this additional constraint is made.}. More specifically, in the first layer, we put $n$ RT surfaces which only surround one elementary region. Then in the second layer, we place the RT surfaces which enclose two adjacent elementary regions, thus we can put more $(n-1)$ RT surfaces. In the third layer, we place the RT surfaces which enclose three adjacent elementary regions, thus we can put more $(n-2)$ RT surfaces, and so on, until we put the last one RT surface enclosing $n$ elementary regions, i.e., the RT surface corresponding to the whole region $A$. In this way, we can always make the two numbers match exactly. From another point of view, we are constructing a one-to-one mapping between the flows and the entropies as follows:
\be\label{map}\begin{array}{l}
	{{\vec v}_{ci}} \to {S_i}\\
	{{\vec v}_{ij}} \to {S_{i\left( {i + 1} \right) \ldots j}}\quad \left( {{\rm{for}}\;i < j} \right)
\end{array} ,\ee
where ${\vec v_{ij}}$ is the component flow connecting two elementary regions ${A_i}$ and ${A_j}$ ($i<j$) within $A$, ${\vec v_{ci}}$ is the component flow connecting ${A_i}$ and ${A_c}$. ${S_i}$ denotes the entanglement entropy of ${A_i}$ region. ${S_{i\left( {i + 1} \right) \ldots j}}$ denotes the entanglement entropy of a composite region ${A_{i\left( {i + 1} \right) \ldots j}} \equiv {A_i} \cup {A_{i + 1}} \cup  \cdots  \cup {A_j}$, which is constructed from several adjacent elementary regions. In particular, we have ${S_{12 \ldots n}} = {S_c}$.

\begin{figure}[htbp]     \begin{center}
		\includegraphics[height=7cm,clip]{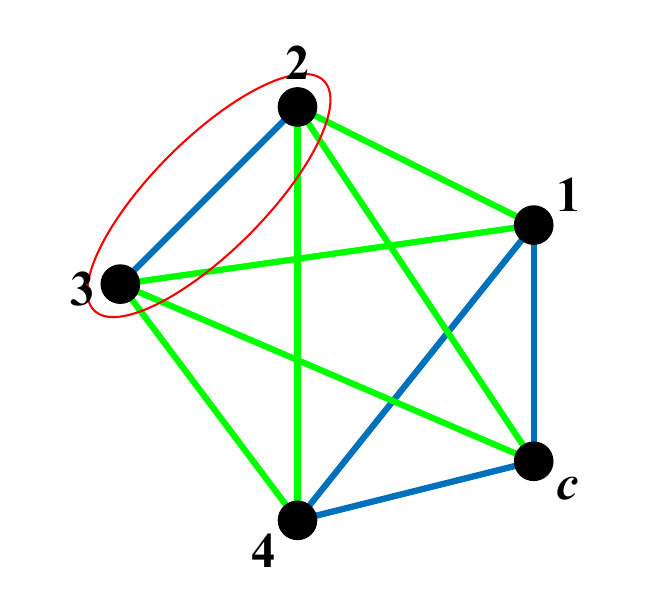}
		\caption{A simplified network diagram corresponding to the figure \ref{fig5}(b). Each elementary region is simplified by a point and the component flow flux is represented by the line segment connecting two points. Then for example, the entropy of the region in red circle is equal to the sum of the green lines in this figure. }
		\label{fig6}
	\end{center}	
\end{figure}

Then we can write a system of equations for a set of unknowns $\left\{ {{F_{ij}},{F_{ci}}} \right\}$ and a set of constraints $\left\{ {{S_i},{S_{i\left( {i + 1} \right) \ldots j}}} \right\}$ of equal numbers, and investigate the solutions. Since we place the RT surfaces in a symmetrical way, we can expect that the system of equations will also present a certain pattern. Indeed, tracing the thread bundles through each RT surface corresponding to the region ${A_{i\left( {i + 1} \right) \ldots j}}$ whose entropy is given by ${S_{i\left( {i + 1} \right) \ldots j}}$, it is not difficult to obtain the following rule:
\be\label{exp}{S_{i\left( {i + 1} \right) \ldots j}} = \sum\limits_{a,b} {{F_{ab}}} \quad {\rm{where}}\;a \in \left\{ {i,i + 1, \cdots ,j} \right\},\;b \notin \left\{ {i,i + 1, \cdots ,j} \right\} ,\ee
in particular, when we take ${{A_{i\left( {i + 1} \right) \ldots j}}}$ as ${A_i}$, we have
\be{S_i} = \sum\limits_{b \ne i} {{F_{ib}}} .\ee
For convenience here $b$ is also allowed to take the subscript $c$ representing the complement of $A$ region, and we adopt the convention that ${F_{ab}} = {F_{ba}}$. This is because for the RT surface corresponding to ${A_{i\left( {i + 1} \right) \ldots j}}$ region, only the threads whose starting point and ending point are inside and outside the ${A_{i\left( {i + 1} \right) \ldots j}}$ region respectively will cross it. The expression~(\ref{exp}) can also be read directly from a simplified network diagram as shown in the figure \ref{fig6}, where each elementary region is simplified by a point and the component flow flux is represented by the line segment connecting two points. When we consider the entropy of the union of a set of adjacent points, we can intuitively circle these points, then the entropy is equal to the sum of the fluxes represented by the line segments whose starting point and ending point are inside and outside the circle respectively. The simplified figure \ref{fig6} also presents the mapping scheme~(\ref{map}) intuitively, the right hand side of ~(\ref{map}) corresponds to the set of circled points, while the left hand side corresponds to the longest line segment connecting the boundary points of the circled part. Now we investigate whether the system of equations~ is solvable. For the case of $n=4$, we may explicitly rewrite it as the matrix equation form as,
\be\left( {\begin{array}{*{20}{c}}
		1&0&0&0&1&1&1&0&0&0\\
		0&1&0&0&1&0&0&1&1&0\\
		0&0&1&0&0&1&0&1&0&1\\
		0&0&0&1&0&0&1&0&1&1\\
		1&1&0&0&0&1&1&1&1&0\\
		0&1&1&0&1&1&0&0&1&1\\
		0&0&1&1&0&1&1&1&1&0\\
		1&1&1&0&0&0&1&0&1&1\\
		0&1&1&1&1&1&1&0&0&0\\
		1&1&1&1&0&0&0&0&0&0
\end{array}} \right)\left[ {\begin{array}{*{20}{c}}
		{{F_{c1}}}\\
		{{F_{c2}}}\\
		{{F_{c3}}}\\
		{{F_{c4}}}\\
		{{F_{12}}}\\
		{{F_{13}}}\\
		{{F_{14}}}\\
		{{F_{23}}}\\
		{{F_{24}}}\\
		{{F_{34}}}
\end{array}} \right] = \left[ {\begin{array}{*{20}{c}}
		{{S_1}}\\
		{{S_2}}\\
		{{S_3}}\\
		{{S_4}}\\
		{{S_{12}}}\\
		{{S_{23}}}\\
		{{S_{34}}}\\
		{{S_{123}}}\\
		{{S_{234}}}\\
		{{S_{1234}}}
\end{array}} \right] .\ee
Again, the determinant of the matrix is non-zero, therefore the matrix has full rank and the system has a unique solution, i.e.,
\be\left[ {\begin{array}{*{20}{c}}
		{{F_{c1}}}\\
		{{F_{c2}}}\\
		{{F_{c3}}}\\
		{{F_{c4}}}\\
		{{F_{12}}}\\
		{{F_{13}}}\\
		{{F_{14}}}\\
		{{F_{23}}}\\
		{{F_{24}}}\\
		{{F_{34}}}
\end{array}} \right] = \left( {\begin{array}{*{20}{c}}
		{\frac{1}{2}}&0&0&0&0&0&0&0&{ - \frac{1}{2}}&{\frac{1}{2}}\\
		{ - \frac{1}{2}}&0&0&0&{\frac{1}{2}}&0&{ - \frac{1}{2}}&0&{\frac{1}{2}}&0\\
		0&0&0&{ - \frac{1}{2}}&{ - \frac{1}{2}}&0&{\frac{1}{2}}&{\frac{1}{2}}&0&0\\
		0&0&0&{\frac{1}{2}}&0&0&0&{ - \frac{1}{2}}&0&{\frac{1}{2}}\\
		{\frac{1}{2}}&{\frac{1}{2}}&0&0&{ - \frac{1}{2}}&0&0&0&0&0\\
		0&{ - \frac{1}{2}}&0&0&{\frac{1}{2}}&{\frac{1}{2}}&0&{ - \frac{1}{2}}&0&0\\
		0&0&0&0&0&{ - \frac{1}{2}}&0&{\frac{1}{2}}&{\frac{1}{2}}&{ - \frac{1}{2}}\\
		0&{\frac{1}{2}}&{\frac{1}{2}}&0&0&{ - \frac{1}{2}}&0&0&0&0\\
		0&0&{ - \frac{1}{2}}&0&0&{\frac{1}{2}}&{\frac{1}{2}}&0&{ - \frac{1}{2}}&0\\
		0&0&{\frac{1}{2}}&{\frac{1}{2}}&0&0&{ - \frac{1}{2}}&0&0&0
\end{array}} \right)\left[ {\begin{array}{*{20}{c}}
		{{S_1}}\\
		{{S_2}}\\
		{{S_3}}\\
		{{S_4}}\\
		{{S_{12}}}\\
		{{S_{23}}}\\
		{{S_{34}}}\\
		{{S_{123}}}\\
		{{S_{234}}}\\
		{{S_{1234}}}
\end{array}} \right] .\ee
It can be verified that according to our interpretation~${s_A}\left( {{A_i}} \right) = {F_{ci}}$, this solution is indeed consistent with the generalized PEE proposal in~\cite{Kudler-Flam:2019oru}, i.e.,
\be\label{gen}{s_A}\left( {{A_i}} \right) = \frac{1}{2}\left[ {S\left( {{A_i}\left| {{A_{1 \cdots \left( {i - 1} \right)}}} \right.} \right) + S\left( {{A_i}\left| {{A_{\left( {i + 1} \right) \cdots n}}} \right.} \right)} \right] ,\ee
where
\be\begin{array}{l}
	S\left( {{A_i}\left| B \right.} \right) = S\left( {{A_i} \cup B} \right) - S\left( B \right)\\
	S\left( {{A_i}\left| {{A_i}} \right.} \right) = S\left( {{A_i}} \right)
\end{array} .\ee
In fact, this is quite easy to understand. The point is that, we can always regard the $n$ elementary regions in $A$ as three regions: ${A_{1'}} = {A_{1 \cdots \left( {i - 1} \right)}}$, ${A_{2'}} = {A_i}$ and ${A_{3'}} = {A_{\left( {i + 1} \right) \cdots n}}$ and thus return to the tripartite case in the previous section. By utilizing the PEE proposal~(\ref{pro}) in tripartite case, we have
\be\label{same}{s_A}\left( {{A_{2'}}} \right) = \frac{1}{2}\left( {{S_{1'2'}} + {S_{2'3'}} - {S_{1'}} - {S_{3'}}} \right) = \frac{1}{2}\left( {{S_{1 \cdots i}} + {S_{i \cdots n}} - {S_{1 \cdots \left( {i - 1} \right)}} - {S_{\left( {i + 1} \right) \cdots n}}} \right) ,\ee
which is exactly the same as (\ref{gen}). On the other hand, it can be seen directly from the figure \ref{fig5} that the locking thread configuration satisfying the requirements of this $n$-partite case should also automatically satisfy the requirement of the tripartite case in the previous. In fact, according to the viewpoint in~\cite{Lin:2020yzf}, the former configuration should be regarded as a more refined description of the entanglement structure of the latter system. Therefore, unsurprisingly, the ${F_{ci}}$ solved from system of the $n$-partite case should satisfy~(\ref{sol}), and thus satisfy~(\ref{same}).

Therefore, it seems that the generalized version of PEE proposal in $n$-partite situations does not say much more than in the tripartite version. Actually, the subtlety is in the conceptual aspect, but not in the technique. Indeed, technically, in order to prove that the PEE of a specified elementary region ${A_i}$ can be obtained by the formula~(\ref{gen}), we can always construct three regions as ${A_{1'}} = {A_{1 \cdots \left( {i - 1} \right)}}$, ${A_{2'}} = {A_i}$, and ${A_{3'}} = {A_{\left( {i + 1} \right) \cdots n}}$, and then directly apply the PEE proposal of tripartite version. However, with this simple method, we are essentially using a locking bit thread configuration which can only guarantee the locking of those specified regions involving in the tripartite case to describe the real entanglement structure of the whole physical system at this tripartite level. At this time, if we check the fluxes ${F_{cj}}$ of other component flows that connecting the other elementary regions ${A_j}$ ($i \ne j$) in $A$ and ${A_c}$, in general, it would not exactly represent the physical meaning of the PEE, and it would not satisfy the generalized formula~(\ref{gen}), since these regions ${A_j}$ have not been locked. Therefore, in order to use our scheme to derive the PEE proposal for the arbitrary $n$-partite situation from the locking thread configuration, what is really important is to ensure that for any value of $n$, the system of equations about ${F_{ci}}$ is solvable, preferably with a unique solution. Because as the increasing of the value of $n$, the order of the matrix associated with the linear system of equations also increases by $\frac{{n\left( {n + 1} \right)}}{2}$. It is not self-evident whether the system involving more and more unknowns ${F_{ci}}$ can be guaranteed to have a unique solution. If the answer is positive, then we can always find the locking thread configuration corresponding to the above scheme, then we can immediately admit the generalized PEE proposal~(\ref{gen}) as we have analyzed. Happily, this is indeed the case. We show that for any value of $n$, the matrix associated with the linear system of equations in our scheme always has full rank, and thus the system always has and only has a unique solution. The main idea of the proof is to use the method of Mathematical induction, and we put the proof in Appendix \ref{app2}.

%%%%%%%%%%%%%%%%%%%%%%%%%%%%%%%%%%%%%%%%%%%%%%%%%%%%%%%%%%%%%%%%%%%%%%
\subsection{Comments on the locking ability of bit threads}\label{subsec4.2}
%%%%%%%%%%%%%%%%%%%%%%%%%%%%%%%%%%%%%%%%%%%%%%%%%%%%%%%%%%%%%%%%%%%%%%

Now, we have to confront a more subtle issue. In essence, to be more precise, we have just proved the existence of a set of thread fluxes $\left\{ {{F_{ci}}} \right\}$ that satisfies our constraint scheme. However, as shown in~\cite{Headrick:2020gyq}, according to the current understanding of bit threads, it seems that the bit threads cannot lock any required set of specified subregions. Because the locking thread configuration not only needs to satisfy the locking constraints of the fluxes, but also to satisfy the nontrivial basic properties of the bit threads per se. For the case of $n=4$, we notice that in~\cite{Headrick:2020gyq}, based on the analogy of the locking problem in the network theory~\cite{net1,net2}, a natural conjecture has been proposed, i.e., a set of regions can be locked if it does not contain a triple of subsets that cross pairwise. It is easy to verify that for the case of $n=4$, there is no pairwise crossing triple in our scheme, so it is probably not too bad. However, when $n$ becomes larger, such pairwise crossing triples will inevitably appear in any locking scheme. For example, when $n=5$, our scheme requires the bit threads to lock ${A_{123}}$, ${A_{234}}$, and ${A_{345}}$ simultaneously, which obviously partially overlap in region ${A_{3}}$.

This predicament can be viewed from two different perspectives. From one perspective, our current judgment on the problem of bit threads is based on the traditional definition of the properties of bit threads. One can imagine that, by a more delicate definition of the properties of bit threads, such as the thread density bound, the ultimate power of bit threads should be able to lock the subregion sets involving any number of regions under any conditions, as long as the constraints of flow fluxes are explicit and reasonable, so as to perfectly adapted to the concept of the PEE. In fact, the limitations of the ability of the current version of bit threads also generally arise in the problem of using the picture of bit threads to prove the higher entropy inequality~\cite{Headrick:2020gyq}, which also requires the bit threads to lock the more general region sets. Based on these reasons, we can think that the coincidence between the idea of PEE and the locking thread configuration is calling for a more delicate definition for bit threads. %%On the other hand, maybe we do not need to make the properties of bit threads more sophisticated, rather, we can take a step back, and simply further relax the density bound in a rude way. For example, we can require that the component flows actually only interact with each other on the involving RT surfaces, while in the bulk they do not affect each other. In other words, the different component flow lives on different ``sheet'', while these sheets are only ``glued'' on the involving RT surfaces and elementary regions. This kind of idea is proposed in~\cite{Headrick:2020gyq,Harper:2019lff}.%%

From another perspective, although the idea of the PEE, i.e., regarding the entanglement entropy of a region as the sum of the contributions of each component part, is natural, but may be too naive. Obviously, the picture of bit threads is likely to be more convincing in realizing this idea of ``the whole equals the sum of its parts'', since bit threads have more nontrivial properties and can precisely reproduce the RT formula mathematically. Therefore, the failure of the locking ability of bit threads in the more refined cases involving large $n$ elementary regions maybe implies that this idea (or belief) is naive to describe the real entanglement structure of a quantum system, especially in the holographic context. As an interesting example, one can compare this idea of ``the whole equals the sum of its parts'' with the idea of the quantum error-correcting code in holography~\cite{Pastawski:2015qua}. In the latter, when we extract some certain information from a region of interest, we are not simply regarding it as the sum of the contributions of each component parts, rather, we regard the whole region as a string of redundant code, and extract the information in a more nontrivial decoding mechanism, which can also be interpreted as the real mechanism of the holographic principle.

%%Nevertheless, our purpose at present is to point out the remarkable coincidence between the locking bit thread configuration and the idea of the PEE. The further implications of this coincidence are beyond the scope of our present investigation. The purpose of the above comments is only to point out that this problem does not necessarily imply an inevitable contradiction.%
Nevertheless, here we would like to provide some immature ideas and intuitions of trying to improve the locking ability of bit threads by modifying their properties. For the density bound 
\be\label{bound1}\rho \left( V \right) \le 1 ,\ee
instead of changing the definition of the thread density on the left hand side, maybe we can try to change its right hand side, for example, as
\be\label{bound2}\rho \left( V \right) \le \alpha \left( x \right) ,\ee
which promotes the thread density bound as a position-dependent parameter. Furthermore, we note that, actually there is a rather radical way that can circumvent the issue of the locking limitations of bit threads. We can require that the component flows actually only interact with each other on the involving RT surfaces and elementary regions, while in the bulk they do not affect each other. In other words, the different component flow lives on different ``sheet'', while these sheets are only ``glued'' on the involving RT surfaces and elementary regions. This idea is hinted at in~\cite{Headrick:2020gyq} and used implicitly in~\cite{ Harper:2019lff}. Then one of the possible directions is to try to quantify this picture into something like~(\ref{bound2}). The clue is that, according to this picture, the thread density contributed by all the thread bundles together should still satisfy~(\ref{bound1}) on the involving RT surfaces and elementary regions, while may exceed $1$ elsewhere in the bulk. Therefore, maybe one can construct some concrete examples to figure out what form the formula~(\ref{bound2}) should take. Moreover, it is also important to re-understand the physical significance of the density bound represented by~(\ref{bound2}). In physics,~(\ref{bound1}) can be naturally understood as requiring that each Planck area (which is assumed to be the smallest physical unit of area ) can accommodate no more than one bit thread. However, a natural physical interpretation of~(\ref{bound2}) is not clear for the moment. Nevertheless, how to improve the locking ability of bit threads is a nontrivial issue, and we will leave the complete solution to this problem for the future.

%%%%%%%%%%%%%%%%%%%%%%%%%%%%%%%%%%%%%%%%%%%%%%%%%%%%%%%%%%%%%%%%%%%%%%
\section{Conclusions and discussions}\label{sec5}
%%%%%%%%%%%%%%%%%%%%%%%%%%%%%%%%%%%%%%%%%%%%%%%%%%%%%%%%%%%%%%%%%%%%%%
In this paper, in the holographic framework, we derive the PEE proposal proposed in~\cite{Wen:2019ubu} and its generalized version in \cite{Kudler-Flam:2019oru} from the picture of bit threads. More specifically, we first explicitly identify the PEE as the flux of the $component~flow$ in a locking bit thread configuration~\cite{Headrick:2020gyq,Lin:2020yzf}. In other words, we identify the entanglement contour as the component flow in a multiflow that describes the locking thread configuration. Then, using the locking theorems of bit threads, we show that when we define the thread density as the number per unit area intersecting a small disk (and maximized over the orientation of the disk), for an arbitrary tripartite subsystem of interest, one can always find a locking thread configuration that can well describe the entanglement structure of the whole system at this level. According to the constraint conditions of this thread configuration, we can uniquely determine the fluxes of the component flows connecting different pairs of elementary regions. It turns out that the expression of the component fluxes obtained in this method is exactly equivalent to the expression of the PEE proposal in~\cite{Wen:2019ubu}. Therefore, in a sense, we have derives the PEE proposal from the picture of bit threads. In other words, we provide a natural and basic interpretation of the PEE proposal in the holographic framework. Moreover, from this perspective of bit threads, we also present a coherent explanation for the coincidence between the BPE (balanced partial entanglement)/EWCS (entanglement wedge cross section) duality proposed recently and the existing EoP (entanglement of purification)/EWCS duality. 

Then we use the language of bit threads to further investigate the PEE proposal in multipartite situations~\cite{Kudler-Flam:2019oru}. We construct a concrete locking scheme for this general situation and prove that under this scheme, the component flow fluxes of the system are uniquely determined and exactly match the formula for the general PEE proposal, given the identification between the component flow flux and the PEE. Although we point out this remarkable coincidence, we have to confront a more subtle issue, i.e., according to the understanding of bit threads at the current stage, it seems that the bit threads cannot lock any required set of specified subregions, because the locking thread configuration should not only satisfy the locking constraints of the fluxes, but also satisfy the nontrivial basic properties of the bit threads per se. In our opinion, on the one hand, this may imply that the bit thread formulation should seek further development so as to have the ability to lock more general region sets. On the other hand, perhaps we should rethink the physical idea of the PEE. At least in the holographic framework, this idea of ``the whole equals the sum of its parts'' may have certain limitations. We will leave it as an open question for the future.

It is worth noting that the locking thread configurations used in this paper only involve the minimal surfaces that grow directly from the boundary. Recently, an interesting so-called ``surface growth scheme'' for reconstructing the holographic bulk geometry was proposed in~\cite{Lin:2020thc,Lin:2020}, in which the minimal surfaces can grow from the more general bulk minimal surfaces. It may be also interesting to investigate the connection between the behavior of the bit threads associated with these more general configurations and the PEE. Another interesting problem is to consider how to generalize our story of the PEE beyond the leading order in ${1/N}$ using the formalism of the quantum bit threads~\cite{Agon:2021tia,Rolph:2021hgz} mentioned in the introduction, which can include the quantum corrections for holographic entanglement entropy. We also leave these questions as future directions.

%%%%%%%%%%%%%%%%%%%%%%%%%%%%%%%%%%%%%%%%%%%%%%%%%%%%%%%%%%%%%%%%%%%%%%
\section*{Acknowledgement}
%%%%%%%%%%%%%%%%%%%%%%%%%%%%%%%%%%%%%%%%%%%%%%%%%%%%%%%%%%%%%%%%%%%%%%
We would like to thank Yuan Sun for useful discussions. This project was supported by the National Natural Science Foundation of China (No.~11675272).

%%%%%%%%%%%%%%%%%%%%%%%%%%%%%%%%%%%%%%%%%%%%%%%%%%%%%%%%%%%%%%%%%%%%%%
\begin{appendix}
%%%%%%%%%%%%%%%%%%%%%%%%%%%%%%%%%%%%%%%%%%%%%%%%%%%%%%%%%%%%%%%%%%%%%%

%%%%%%%%%%%%%%%%%%%%%%%%%%%%%%%%%%%%%%%%%%%%%%%%%%%%%%%%%%%%%%%%%%%%%%
\section{The proof by Mathematical induction}\label{app2}
The proof is using Mathematical induction.

\begin{figure}[htbp]     \begin{center}
		\includegraphics[height=6.7cm,clip]{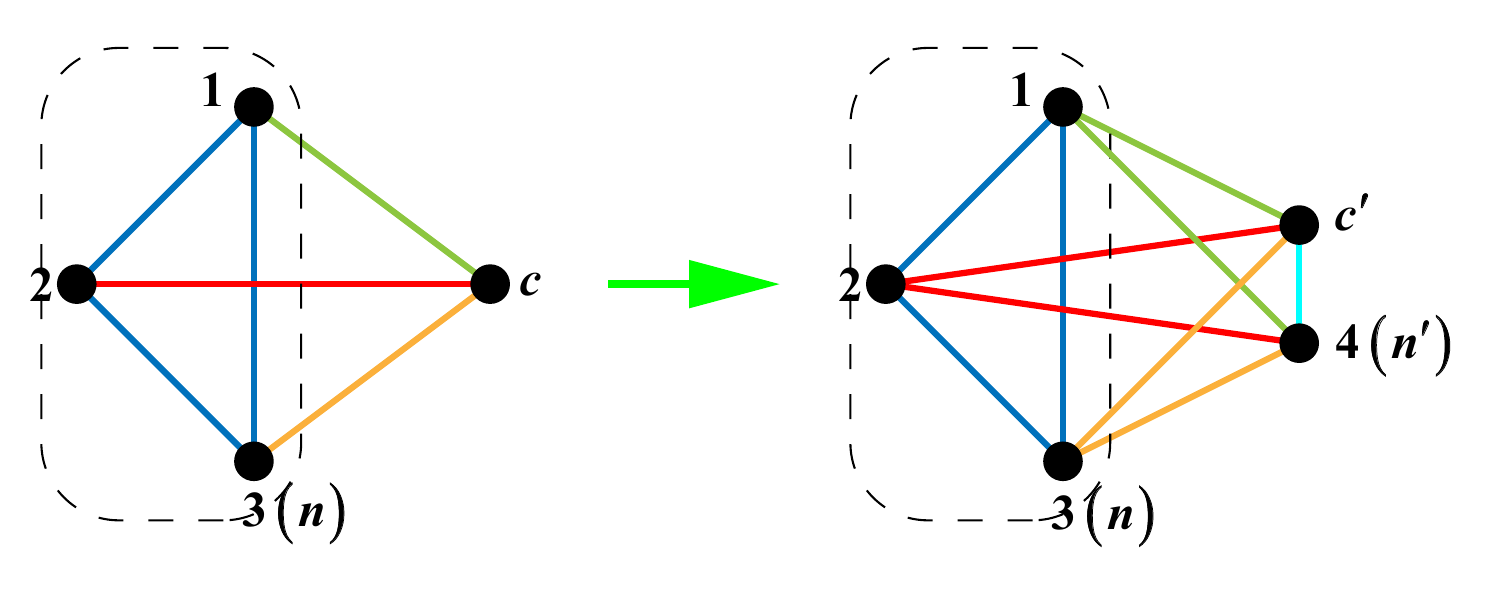}
		\caption{When ${A_c}$ is further decomposed into two new elementary regions ${A_{c'}}$ and ${A_{n'}}$, the $n$ number of original flow fluxes ${F_{ci}}$ is now replaced by $2n$ number of flow fluxes $\left\{ {{F_{c'i}},{F_{n'i}}} \right\}$, and a new flow flux ${F_{c'n'}}$ is added. We use different colors to show the changing of the flows in this process.}
		\label{fig7}
	\end{center}	
\end{figure}

As shown in the figure \ref{fig7}, supposed we have found a set of flow fluxes $\left\{ {F_{^{ij}}^{\left( 0 \right)},F_{ci}^{\left( 0 \right)}} \right\}$ describing a locking thread configuration, which is the unique solution of the system of equations $\mathscr{E}$ involving $(n+1)$ elementary regions $\left\{ {{A_c},{A_1},{A_2}, \cdots ,{A_n}} \right\}$ and the required composite regions $\left\{ {{A_{i\left( {i + 1} \right) \ldots j}}} \right\}$ in our locking scheme (From here on, we will take $i$ or $j$ as $1$ to $n$, but not $c$, and $c'$, $n'$ below). But now we further divide ${A_c}$ into two elementary regions, denoted as ${A_{c'}}$ and ${A_{n'}}$ respectively, so we have $(n+2)$ elementary regions $\left\{ {{A_{c'}},{A_1},{A_2}, \cdots ,{A_n},{A_{n'}}} \right\}$. Then we can ask, is there still a unique solution to the linear system of equations $\mathscr{E'}$ corresponding to this new situation under our scheme? We will show that the answer is positive.

As can be seen intuitively from the network figure \ref{fig7}, when ${A_c}$ is further decomposed into two new elementary regions ${A_{c'}}$ and ${A_{n'}}$, what happens physically is that the $n$ number of original flow fluxes ${F_{ci}}$ is now replaced by $2n$ number of flow fluxes $\left\{ {{F_{c'i}},{F_{n'i}}} \right\}$, and a new flow flux ${F_{c'n'}}$ is added. Therefore, the number of unknowns in the system increases by $(n+1)$, from $\frac{{n\left( {n + 1} \right)}}{2}$ to $\frac{{\left( {n + 2} \right)\left( {n + 1} \right)}}{2}$. The key point is to notice that essentially physically the thread configuration represented by the solution of the new system $\mathscr{E'}$ should still describe the same entanglement inside the $A$ region (i.e., ${F_{ij}}$) and the entanglement between $A$ and its complement $ {A_c}$ (i.e., ${F_{ci}}$), expect that this new thread configuration is more refined than the old one, and thus can also describe the entanglement between $A$ and the two different parts of $ {A_c}$ (i.e., $\left\{ {{F_{c'i}},{F_{n'i}}} \right\}$) in more details and the entanglement between ${A_{c'}}$ and ${A_{n'}}$ within the $ {A_c}$ region (i.e., ${F_{c'n'}}$). Furthermore, once we find a unique solution to the old system $\mathscr{E}$, the $\frac{{n\left( {n + 1} \right)}}{2}$ number of unknows become the known data in terms of the entropies of the specified regions. Therefore, in the new system $\mathscr{E'}$, the value of the unknows ${F_{ij}}$ must be fixed as the same as the solution $F_{^{ij}}^{\left( 0 \right)}$ of the old system $\mathscr{E}$. This is intuitively represented in the simplified network figure \ref{fig7}, where we require the line segments enclosed in the circle do not change. Moreover, the value of ${F_{ci}} = {F_{c'i}} + {F_{n'i}}$ is fixed as $F_{ci}^{\left( 0 \right)}$. Therefore, in solving the new system $\mathscr{E'}$, actually we are only dealing with $(n+1)$ unknows, including $n$ number of ${F_{n'i}}$ (while ${F_{c'i}}$ can be directly represented as ${F_{c'i}} = F_{ci}^{\left( 0 \right)} - {F_{n'i}}$) and one ${F_{n'c'}}$, just as shown in figure \ref{fig7}, the final graph increases $(n+1)$ segments starting from $n'$ point compared to the original graph. One the other hand, we also obtain exactly $(n+1)$ new physical constraints $\left\{ {{S_{n'}},{S_{nn'}},{S_{\left( {n - 1} \right)nn'}},{S_{1 \cdots nn'}}} \right\}$ which correspond to the different ways of circling a set of adjacent points starting from $n'$. Therefore, we can introduce $(n+1)$ number of equations for these $(n+1)$ number of unknows, according to~(\ref{exp}), we have
\be\begin{array}{l}
	{S_{n'}} = \sum\limits_{j \in \left\{ {1,2, \ldots ,n} \right\}} {{F_{n'j}}}  + {F_{n'c'}}\\
	{S_{nn'}} = \left( {\sum\limits_{j \in \left\{ {1,2, \ldots ,n - 1} \right\}} {F_{nj}^{\left( 0 \right)}}  + {F_{nc'}}} \right) + \left( {\sum\limits_{j \in \left\{ {1,2, \ldots ,n - 1} \right\}} {{F_{n'j}}}  + {F_{n'c'}}} \right)\\
	\cdots  \cdots \\
	{S_{i(i + 1) \ldots nn'}} = \left( {\sum\limits_{j \in \left\{ {1,2, \ldots ,i - 1} \right\}} {F_{ij}^{\left( 0 \right)}}  + {F_{ic'}}} \right) + \left( {\sum\limits_{j \in \left\{ {1,2, \ldots ,i - 1} \right\}} {F_{\left( {i + 1} \right)j}^{\left( 0 \right)}}  + {F_{\left( {i + 1} \right)c'}}} \right)\\
	\quad \quad \quad \quad \quad\quad\quad  +  \cdots  + \left( {\sum\limits_{j \in \left\{ {1,2, \ldots ,i - 1} \right\}} {F_{nj}^{\left( 0 \right)}}  + {F_{nc'}}} \right) + \left( {\sum\limits_{j \in \left\{ {1,2, \ldots ,i - 1} \right\}} {{F_{n'j}}}  + {F_{n'c'}}} \right)\\
	\cdots  \cdots \\
	{S_{12 \ldots nn'}} = {F_{1c'}} + {F_{2c'}} +  \cdots  + {F_{nc'}} + {F_{n'c'}}
\end{array} .\ee

Now notice that we can substitute ${F_{ic'}} = F_{ci}^{\left( 0 \right)} - {F_{n'i}}$ into the system, and since the quantities with a $(0)$ superscript are known constants, we can move all these constants into the left-hand side, such that there are only unknowns ${F_{n'i}}$ and ${F_{n'c'}}$ in the right-hand side. Redefining the left-hand side as $\tilde S$, we thus have
\be\begin{array}{l}
	{{\tilde S}_{n'}} = {F_{n'c'}} + {F_{n'1}} + {F_{n'2}} +  \cdots  + {F_{n'\left( {n - 1} \right)}} + {F_{n'n}}\\
	{{\tilde S}_{nn'}} = {F_{n'c'}} + {F_{n'1}} + {F_{n'2}} +  \cdots  + {F_{n'\left( {n - 1} \right)}} - {F_{n'n}}\\
	\cdots  \cdots \\
	{{\tilde S}_{i(i + 1) \ldots nn'}} = \left( {{F_{n'c'}} + {F_{n'1}} + {F_{n'2}} +  \cdots  + {F_{n'\left( {i - 1} \right)}}} \right) - \left( {{F_{n'i}} + {F_{n'\left( {i + 1} \right)}} +  \cdots {\rm{ + }}{F_{n'n}}} \right)\\
	\cdots  \cdots \\
	{{\tilde S}_{12 \ldots nn'}} = {F_{n'c'}} - \left( {{F_{n'1}} + {F_{n'2}} +  \cdots  + {F_{n'n}}} \right)
\end{array} ,\ee
which can be written as the matrix equation
\be\left[ {\begin{array}{*{20}{c}}
		{{{\tilde S}_{n'}}}\\
		{{{\tilde S}_{nn'}}}\\
		{{{\tilde S}_{\left( {n - 1} \right)nn'}}}\\
		\vdots \\
		\vdots \\
		{{{\tilde S}_{2 \ldots nn'}}}\\
		{{{\tilde S}_{12 \ldots nn'}}}
\end{array}} \right] = \left( {\begin{array}{*{20}{c}}
		1&1&1& \cdots &1&1&1\\
		1&1&1& \cdots &1&1&{ - 1}\\
		1&1&1& \cdots &1&{ - 1}&{ - 1}\\
		\vdots & \vdots & \vdots & \vdots & \vdots & \vdots & \vdots \\
		1&1&1&{ - 1}&{ - 1}& \cdots &{ - 1}\\
		1&1&{ - 1}&{ - 1}&{ - 1}& \cdots &{ - 1}\\
		1&{ - 1}&{ - 1}&{ - 1}&{ - 1}& \cdots &{ - 1}
\end{array}} \right)\left[ {\begin{array}{*{20}{c}}
		{{F_{n'c'}}}\\
		{{F_{n'1}}}\\
		{{F_{n'2}}}\\
		\vdots \\
		\vdots \\
		{{F_{n'\left( {n - 1} \right)}}}\\
		{{F_{n'n}}}
\end{array}} \right] .\ee 
It can be seen that the matrix associated with the system has a very nice pattern, in which the elements on the diagonal and the upper left side of the diagonal are all $1$, while the elements on the lower right side of the diagonal are all $-1$. We can immediately prove that this square matrix has full rank, because we can perform a special elementary transformation on each row of the matrix: add the first row to each row and divide by $2$. Then the matrix becomes
\be\left( {\begin{array}{*{20}{c}}
		1&1&1& \cdots &1&1&1\\
		1&1&1& \cdots &1&1&0\\
		1&1&1& \cdots &1&0&0\\
		\vdots & \vdots & \vdots & \vdots & \vdots & \vdots & \vdots \\
		1&1&1&0&0& \cdots &0\\
		1&1&0&0&0& \cdots &0\\
		1&0&0&0&0& \cdots &0
\end{array}} \right) ,\ee
which obviously has full rank. According to the basic knowledge of linear algebra, an elementary transformation does not change the rank of a matrix. Therefore, the original matrix also has full rank, the system thus has a unique solution.

In conclusion, we have proved that under our scheme, if there exists a unique solution for a set of flow fluxes that can simultaneously lock the region set $I = \left\{ {{A_i},{A_{i\left( {i + 1} \right) \ldots j}}\left| {i,j \in \left\{ {1,2, \ldots ,n} \right\}} \right.} \right\}$ involving $(n+1)$ number of elementary regions $\left\{ {{A_c},{A_1},{A_2}, \cdots ,{A_n}} \right\}$, then there always exists a unique solution for a set of flow fluxes that can simultaneously lock the region set $I = \left\{ {{A_i},{A_{i\left( {i + 1} \right) \ldots j}}\left| {i,j \in \left\{ {1,2, \ldots ,n+1} \right\}} \right.} \right\}$ involving $(n+2)$ number of elementary regions $\left\{ {{A_c},{A_1},{A_2}, \cdots ,{A_{n+1}}} \right\}$. Since we have shown in the case of $n=2$ there exists the unique solution, by Mathematical induction, for any case of $n$, there always exists the unique solution.

\end{appendix}

%%%%%%%%%%%%%%%%%%%%%%%%%%%%%%%%%%%%%%%%%%%%%%%%%%%%%%%%%%%%%%%%%%%%%%

%\end{thebibliography}

\end{document}